\documentclass[]{aa}
\usepackage{graphicx}
\usepackage{natbib}
\usepackage{longtable}

\begin{document}
\title{Chemically peculiar stars and their temperature calibration 
\thanks{}}
\author{M.~Netopil\inst{1}, E.~Paunzen\inst{1}, H.M.~Maitzen\inst{1}, P.~North\inst{2}, S.~Hubrig\inst{3} }

\mail{martin.netopil@univie.ac.at}

\institute{Institut f\"ur Astronomie der Universit\"at Wien,
           T\"urkenschanzstr. 17, A-1180 Wien, Austria		 
\and Laboratoire d'Astrophysique, Ecole Polytechnique F\'ed\'erale de Lausanne (EPFL), Observatoire de Sauverny, CH-1290 Versoix, Switzerland
\and European Southern Observatory, Casilla 19001, Santiago, Chile
}

\date{Received 2008; Accepted 2008}
\authorrunning{M. Netopil et al.}{}
\titlerunning{Chemically peculiar stars and their temperature calibration}{}

\abstract{}
{The determination of effective temperature for chemically peculiar (CP) stars by means of photometry is a sophisticated task due to their abnormal colours. Standard calibrations for normal stars lead to erroneous results and, in most cases corrections are necessary.}
{In order to specify appropriate corrections, direct temperature determinations for 176 objects of the different subgroups were collected from the literature. This much larger sample than in previous studies therefore allows a more accurate investigation, mostly based on average temperatures.}
{For the three main photometric systems ($UB\it{V}, Geneva$, Str\"omgren $uvby\beta$), methods to determine effective temperature are presented together with a comparison with former results. Based on the compiled data we provide evidence that He (CP4) objects also need a considerable correction, not noticed in former investigations due to their small number. Additionally, a new relation for the bolometric correction and the capability of standard calibrations to deduce interstellar reddening for magnetic CP stars are shown.}
{}


\maketitle

\section{Introduction}

There are only a few direct temperature determinations available of chemically peculiar (CP) stars (results other than from photometry), insufficient to study their evolutionary status. Additionally, photometric standard calibrations for normal stars are mostly inappropriate because of their anomalous properties, e.g. blanketing effects, individual abundances or magnetic fields influencing the photometric colours. Using some peculiar stars with direct temperature determinations, $Geneva$ and Str\"omgren photometry was recalibrated by \citet{HN93}, \citet{NA93}, \citet{S94}, or \citet{HK96}. However, small numbers have often allowed only a vague estimate of the CP stars temperature behaviour, best seen on the basis of the CP3 (HgMn) sample used by \citet{HN93} comprising only five objects. \citet{NA93} have included only two stars for comparison, but for the group of CP4 stars insufficient data also were available. Several new temperature determinations having been published, we decided to refine the available corrections and calibrations for the abovementioned  photometric systems based on a larger sample. Due to the fact that numerous new CP2 (magnetic group with enhanced Si, Cr, Sr or Eu) objects also have been detected in galactic open clusters or even in the Large Magellanic Cloud (e.g. \citealt{N07}; \citealt{PMP06}), and photometric data in the abovementioned systems are scarce at larger distances, an investigation of the widely used $UB\it{V}$ photometric system is necessary in order to examine the (extra)galactic distribution of CP objects in detail. 

\section{Data collection}
\label{collection}
Our used starting point was the compilation of chemically peculiar ``standards'' in previous temperature calibration investigations (e.g. \citealt{HN93}). Due to the small number of available direct temperature determinations, most of the same stars have been used for the calibration of the different photometric systems. We consulted the literature to collect more temperatures reported to date and older ones ignored in previous compilations and included only temperature determinations not based on photometry. Results based on averaged values including photometric results were rejected, the same holds for works with no clear description of the method used. Furthermore, the objects were checked for membership of one of the CP groups following the classification scheme by \citet{P74} primarily using the peculiarity types given in \citet{R91} and \citet{BBM03}, refined with additional literature values. The He representatives are divided into CP4a (He-weak) and CP4b (He-rich). For the CP3 members subgroups also have been defined (see Sect. \ref{cp4stars}). If a classification was uncertain, the object was rejected. For instance, the star HD~2628 was found to be a nonclassical Am star by \citet{ACK00}. An exception was made in the case of the stars HD~5737 and HD~125823. \citet{HG99} classified them as intermediate stars which show He-weak and He-rich behaviour, they were kept as CP4ab. For two cool CP2 stars (HD~29578 and HD~92499, both with $T_\mathrm{eff} < 8000$) we are unable to calibrate their temperature, since only $Geneva$ photometry was available and therefore it was not possible to deduce reddening information. Since for these two stars only one temperature determination was found, they were rejected from the sample. In total, 364 individual temperature determinations for 176 CP stars taken from 71 references have been found that fulfil the criteria above. Since we have not noticed significant differences in the overall result between the various methods, no weighting was performed. Some outliers are found, but a rejection was only carried out if more than two determinations were available and a temperature by the same or comparable method was deviated strongly. This was necessary for only about 2\% of the nearly 400 individual results. Finally an average and the resulting standard deviation was calculated. For about 92\% of the sample the standard deviation does not exceed 5\%. Some outliers are based on two references only, or the literature values are spread equally over a large temperature range, making a rejection of individual results difficult. 

Photometric data in the studied systems ($UB\it{V}$, Str\"omgren $uvby\beta$, $Geneva$) were collected by using the \textit{General Catalogue of Photometric Data} (GCPD, \citealt{MMH97}). Since the averaged photometry is based on many references with several individual measurements, in addition to wrong photometry, the influence of several kinds of variability is reduced. If no data were found, the literature was consulted in order not to ignore existing measurements. An overview of the number of CP stars compiled can be seen in Table \ref{sample}. 

The compiled references were divided into groups of comparable temperature determination methods. These are (a) the Infrared Flux Method (IRFM) introduced by \cite{BS77}, (b) fitting solar/enhanced models to the visual energy distribution, (c) fitting solar/enhanced models to the total energy distribution (from UV to red) as well as (d) fitting (Balmer) line profiles with solar/enhanced models. Three references (e) do not fit into these categories. \cite{SD89} presented a new method based on visual energy distribution modelling and a correction of the UV flux deficit. \cite{CAT07} investigated He peculiar stars using a spectrophotometric system based on the measurement of the continuum energy distribution around the Balmer discontinuity. Since this method was not applicable to their He-rich subsample, non-LTE model atmospheres were used for this CP group. \cite{BNC08} determined for the first time the temperature of HD~128898 using angular diameter and bolometric flux. In Table \ref{cpref} we present the references for the individual CP stars together with the method used, (a)$-$(e). This is a rough division, but can serve as a hint to reliability of a particular determination. Falling into group (d) one can find works like \cite{HN07}, who used iterative processes to minimise the dependence of the average Fe abundance on the excitation potentials of several measured lines, but also several works like \cite{RLK04} using a single line such as H$\alpha$. However, further subdivision will probably result in a confusing number of groups. Several references used combinations of these methods, e.g. in the series by Adelman and collaborators often visual energy distribution modelling plus H$\gamma$ profile fitting were used, which is indicated in Table \ref{cpref} as bd(+), where the plus sign shows that a model other than a solar one was used. Several objects in the comprehensive list by \cite{AR00} have been studied in previous works by Adelman with good agreement. Since it seems that the same data have been used, we only include the results of the latter reference. Furthermore, \cite{AR00} used more recent model atmospheres than in the previous studies. 

The mean effective temperatures of the compiled CP stars (Tables \ref{cp1list}$-$\ref{cp4list}) are therefore based on several individual studies, determined mostly with different methods (see Table \ref{cpref}) compensating for the possible disadvantages of a particular method.
\begin{table}
\begin{minipage}[t]{85mm}
\caption{The sample of CP stars used in this study and the available photometric data in the respective systems.} 
\label{sample} 
\centering 
\renewcommand{\footnoterule}{}
\begin{tabular}{l l l l l} 
\hline\hline 
CP Class & Stars\footnote{Stars in total / with average temperatures} & $uvby\beta$ & $Geneva$ & $UB\it{V}$\footnote{Number of objects with a complete set is listed}  \\ 
\hline 
CP1 & 30/13  & 29 & 30 & 30 \\
CP2 & 79/51  & 78 & 76 & 63  \\
CP3 & 28/14  & 27 & 27 & 28  \\
CP4a & 20/15  & 20 & 20 & 20 \\
CP4b & 17/9 & 14 & 16 & 15 \\
CP4ab & 2/2 & 2 & 2 & 2 \\
\hline 
\end{tabular}
\end{minipage}
\end{table}  

\section{Interstellar reddening}
\label{reddening}
\begin{figure}[t]
\begin{center}
\includegraphics[width=80mm]{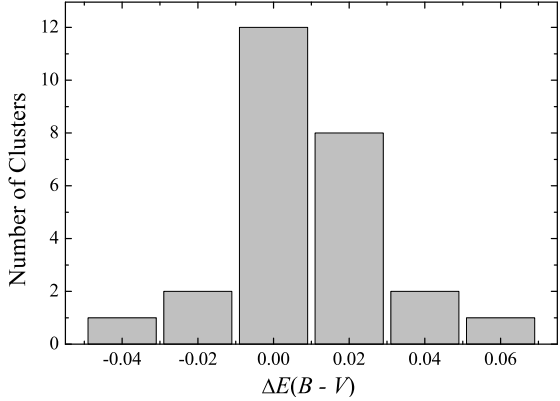}
\caption{Histogram of the mean reddening deviations $\Delta E(B-V)$.}
\label{cpebv}
\end{center}
\end{figure}

For the vast majority of programme stars Hipparcos parallaxes (\citealt{PLK97}) are available. A new reduction of the data (\citealt{vL07}) has been published, providing more accurate results. These are used for the present study. Since only a limited number of objects was found to be located closer than 50\,pc from the sun, interstellar extinction is no longer negligible, especially if examining cooler CP stars in the $Geneva$ or $UB\it{V}$ photometric system, for which a reddening-free temperature calibration (via $Geneva$ $X/Y$ or the $UB\it{V}$ $Q$ method) is not possible. Several attempts have been made to now to model the distribution of interstellar extinction. However, these studies should be treated with caution if one intends to deredden individual stars, because such models give only a general trend and do not take local irregularities of the absorbing material into account (\citealt{AGG92}). We therefore rely on reddening estimations based on $UB\it{V}$, $Geneva$ and $uvby\beta$ data. To examine the applicability and accuracy for CP stars because of their anomalous colours, we have chosen the compilation of magnetic CP stars in open clusters by \citet{LB07}. The available listing was reduced by using their flags to limit to at least probable cluster CP objects. 
We also removed objects within associations and young cluster stars ($\log t \le 7.0$) to avoid strong differential reddening due to nebulous regions. Averaged cluster ages for the selection and the reddening values for the comparison are taken from \citet{PN06}. For clusters not included in their list, we have proceeded analogously. For the remaining 45 CPs in 26 open clusters that have a reddening up to 0.5\,mag, we have extracted photometric measurements in the three mentioned systems using the GCPD and the literature. For all objects, data in at least one system are available.
Using the \citet{NA93} UVBYBETA calibration, the intrinsic $Geneva$ colours (\citealt{CRAM82}) via the $X/Y$ parameters and the $Q$ method for the $UB\it{V}$ system (\citealt{J58}), $$E(B-V)=(B-V)-0.332Q$$ $$Q=(U-B)-0.72(B-V)-0.05(B-V)^2$$ the colour excesses in the respective systems are determined. However, the latter two systems can be used only for hotter stars. Following the suggestion by \citet{N98}, the correction for hotter CP2 stars in the $uvby\beta$ system as defined by \citet{MJM98} was applied to take the peculiarity effects on $c_{1}$ and $(b-y)$ into account. Using this method, negative reddening values are reduced reasonably; the remaining ones are set to zero. Finally, the relations $$E(B-V) = 1.43E(b-y) = 0.84E[B-V]$$ are used to transform reddening values of the different photometric systems to calculate a mean reddening. Square brackets are utilised to distinguish the $Geneva$ excess from Johnson $E(B-V)$. Note the transformations $E(B-V)=1.14E(B2-V1)=0.83E(B2-G)$ for the additional $Geneva$ colour excesses. To reduce the influence of differential reddening, which is significantly present in NGC~2516 (see e.g. \citealt{MH81}), the individual determined colour excesses for CP stars $E(B-V)_{CP}$ in a cluster are averaged whenever possible. Figure \ref{cpebv} shows the resulting deviations from the mean cluster reddening ($\Delta E(B-V) = E(B-V)_{Cluster} - E(B-V)_{CP}$). The CP star HD~127924 in NGC~5662 exhibits the largest difference from the mean cluster reddening (0.06\,mag). When inspecting the reddening distribution by \citet{CLB91} in this cluster, it is obvious that it lies in a slightly less reddened region. 

The mean deviation was found to be comparable to the errors of the average cluster reddenings ($\sim$\,0.02\,mag). We therefore conclude that the use of the photometrically determined colour excesses is justified and it was applied to the magnetic groups of our sample. However, for objects closer than 50\,pc we still assume non reddening. The methods above cannot be used safely in regions with an exotic reddening law.

\section{Photometric temperature calibration}

For all stars with available Str\"omgren $uvby\beta$ photometry, their initial temperature was obtained using the UVBYBETA calibration by \citet{NA93}, hereafter denoted $T_{uvby\beta}$. For the hotter stars with $Geneva$ photometry, the temperature ($T_{XY}$) was estimated using the reddening-free $X/Y$ parameters and the calibration by \citet{KNK97}. For both systems the grids for $\mathrm{[M/H]}=0$ are used, in order to have the possibility of a direct comparison to former studies. The stars calibratable via $X/Y$ parameters can be selected by using $X \la 1.7+3Y$ and $Y \ga  -0.07$ (\citealt{HN93}) as an approximation. For this subsample the $UB\it{V}$ $Q$-parameter also can be calculated safely. Whenever possible, the best suited relation based on the Str\"omgren reddening-free indices is given, which is in all cases the $[u-b]$ index.

One crucial point to determine the corrections for the individual CP subgroups and photometric systems 
is the consideration of errors. \citet{NA93}, \citet{S94} or \citet{LS08} have not used errors for the calculations of their $[u-b]$ and $[c_{1}]$ relations, in contrast to \citet{HN93} who assumed 300\,K in their $Geneva$ correction for both axes. Inspecting their results for CP2 stars with and without the use of errors, a non-significant difference can be found. Since the ``real'' errors are unknown, and even the standard deviations of the averaged temperatures cannot be considered as realistic errors, we omitted them.

In Table \ref{relations} an overview of all determined relations and the respective errors, their validity range and the correlation coefficient $R$ can be found. In the following sections the individual CP groups are discussed together with a comparison to former results.

\subsection{CP1}
For the group of Am stars the situation is straightforward, since well known calibrations for 
normal stars can be used with high accuracy. Although they are metal-rich, solar composition within the UVBYBETA calibration provides the best results ($\Delta T < 200$\,K) without the need for any correction. Using the more realistic assumption of [M/H]=0.5, the temperature is systematically underestimated by $\sim$150\,K. In the light of the findings by \citet{HN93} that unrecognized binarity (most of the Am stars are SB1 objects) lowers the apparent effective temperature by 2.5-3\%, it seems that using a solar composition grid for the photometric calibration balances this effect. However, in the literature it was not always traceable whether binarity was taken into account for the determined effective temperatures.

In contrast to $uvby\beta$ photometry, interstellar reddening has to be known if examining cool stars with $Geneva$ photometry. Except for one star (HD~162132), all objects of our sample are closer than 100\,pc, about one third are even members of the nearby non-reddened open cluster Hyades. Examining the reddening determined via $uvby\beta$ photometry, one can notice that only three stars exhibit an $E(B-V)$ greater than 0.02\,mag. In consideration of these facts and that the mean reddening of the sample is 0.004$\pm$0.008\,mag, we have omitted a reddening correction (also for $UB\it{V}$), evoking an error of 225\,K for the hottest part assuming an error of 0.02\,mag in reddening. This is just slightly higher than the mean standard deviation of the individual average effective temperatures (150\,K). However, inspecting the reddening determinations via $uvby\beta$ for the Hyades stars an excellent agreement with the mean cluster reddening was found.

The easiest and most accurate way to determine effective temperatures via $Geneva$ photometry is the use of the $(B2-V1)_{0}$ relation given in \citet{HAU85}: $$\theta_\mathrm{eff}=0.632+0.640(B2-V1)_{0}$$ for $-0.160 \leq (B2-V1)_{0} \leq +0.730$ ($\theta_\mathrm{eff}=5040/T_\mathrm{eff}$). See also \citet{HN93} in this respect. \citet{HK96} have proposed, beside the $(B2-V1)$ relation above, also the direct use of the calibration by \citet{KNK97}, but the resulting temperatures are consistently underestimated by about 200\,K for stars cooler than $\sim$9000\,K and by the same value too high for the hotter part. 

For stars in the temperature domain where the $Geneva$ reddening-free $X/Y$ parameters can be used directly, no correction of the calibration by \citet{KNK97} is necessary. However, the sample of such hot Am representatives is rather small. 

In the case of Johnson $UB\it{V}$, a relation based on $(B-V)$ colours for normal stars (e.g.\citealt{FLO96}) results in temperatures about 200\,K too low, caused by line blanketing due to metallic lines (\citealt{FEI74}), that has no influence on the $Geneva$ $(B2-V1)$ index (\citealt{HN93}). To obtain a proper calibration, objects deviating more than 150\,K after applying the $(B2-V1)$ relation are rejected to reduce effects of interstellar reddening or other individual variances.  

\begin{figure}[t]
\begin{center}
\includegraphics[width=80mm]{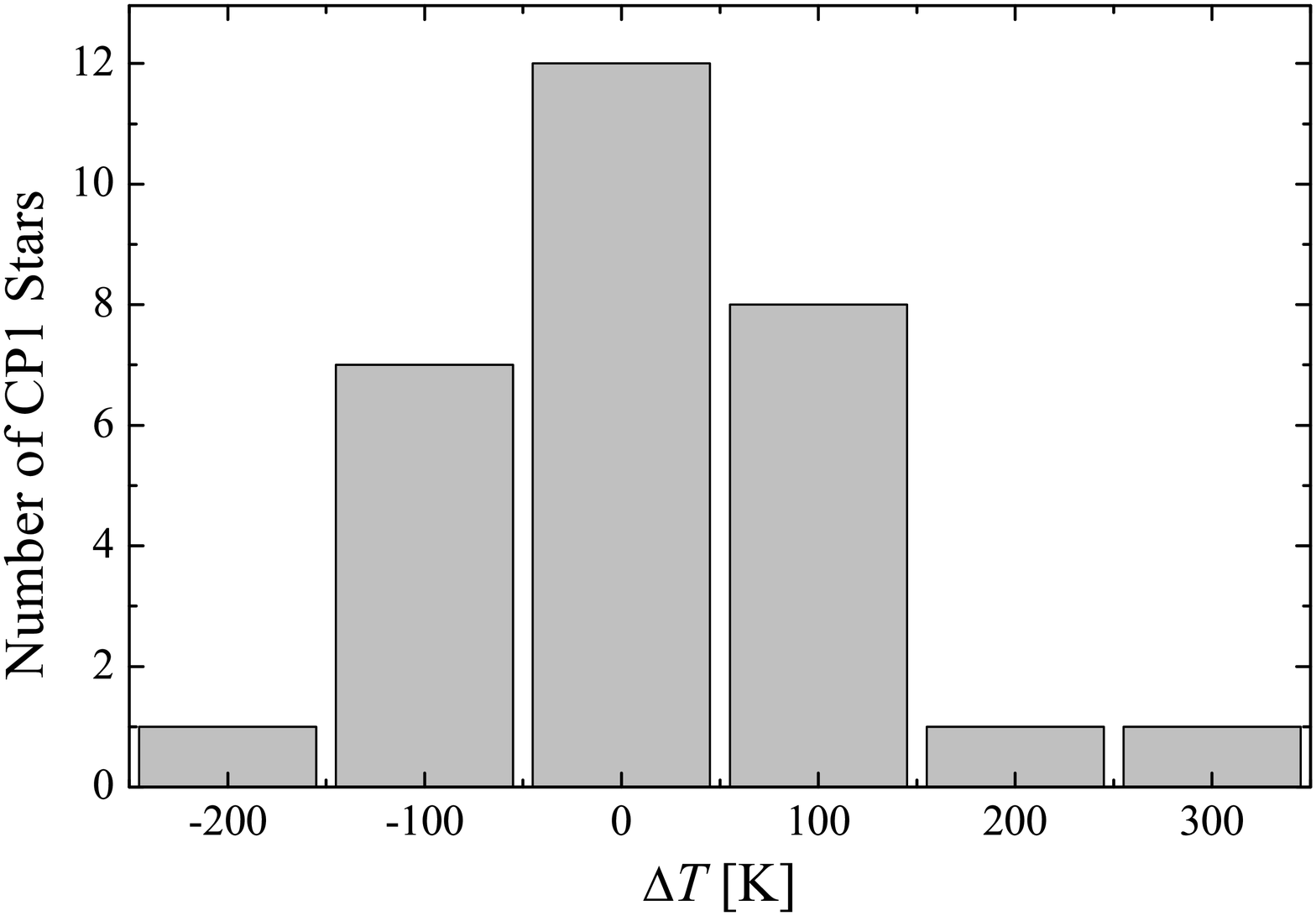}
\caption{Histogram of the temperature deviations for CP1 stars ($T_\mathrm{eff} - \overline{T}_{phot}$).}
\label{Amtemp}
\end{center}
\end{figure}

All investigated systems can be used to calibrate effective temperatures at about the same accuracy level 
($\sim$ 200\,K). However, $uvby\beta$ photometry should be preferred because of the possibility to deduce 
interstellar reddening. If the colour excess is known or the stars are close by, an average of the three 
systems results in high precision. Figure \ref{Amtemp} shows the histogram of the deviations ($\Delta T = T_\mathrm{eff}-\overline{T}_\mathrm{phot}$); only three stars exhibit a deviation of more than 150\,K, whereas $\sim$80\% are calibrated to better than 100\,K.

\subsection{CP2}
\label{lipski}
Several studies in the past dealt with the temperature calibration for the CP2 group, the most recent being by \citet{LS08}. Since our sample for this peculiar type is rather large (79 objects) compared to previous studies, we are able to restrict it to a selection of the most accurate stars, such CP2s with at least two independent temperature determinations. These 51 objects still cover the complete temperature range of 7000-14000\,K, including several cool roAP stars. A lack of such cool representatives occured in previous attempts. To obtain a more realistic error analysis of the different corrections, these are applied to the whole sample in order to take the different properties and error sources of CP2 stars into account.

Within this limited sample, one star (HD~133880) shows a large discrepancy if comparing effective temperatures and temperatures determined via standard photometric calibrations, and was excluded from the analysis. \citet{S94} argued that one can find good reasons to reject almost every peculiar star from a sample. However, this star is an exceptional one due to the strong magnetic field and its geometry (\citealt{L90}). 

One additional object deserves closer attention. HD~173650 was investigated by two authors (\citealt{W67}; \citealt{BAB94}), but with different results; they obtained 9000 and 11000\,K, respectively. Since the average value does not affect the correlations in all photometric systems, we decided to keep it in our sample.

In case of the UVBYBETA calibration, stars resulting in temperatures $T_{uvby\beta}<9000$\,K can be used without correction. For hotter stars a correction is necessary, listed in Table \ref{relations}. If both cases are applied properly to the whole sample, an accuracy better than $\sim$500\,K can be achieved. Some outliers are present, but it is not possible to distinguish whether they are due to a wrong effective temperature determination or because of individual anomalies. 

\citet{AR00} proposed a temperature correction for CP2 stars based on results of the UVBYBETA calibration by \citet{NA93}. They found $T_{uvby\beta} = 1.1984T(sp)-1704$ compared to their spectrophotometrically determined temperatures using 17 stars. Applied to our sample, the cool part ($\la$ 11000\,K) is reproduced suitably, whereas the hotter stars are overestimated by about 500\,K. This can be explained by the large scatter in the results among the hotter ones.

Recently, \citet{LS08} presented revised calibrations by means of the reddening free $[u-b]$ and $[c_{1}]$ indices. They found that a quadratic fit is necessary to take the CP2 properties into account. Based on our sample we cannot confirm these results, although their determined temperatures are included in our sample.

\begin{figure}[t]
\begin{center}
\includegraphics[width=80mm]{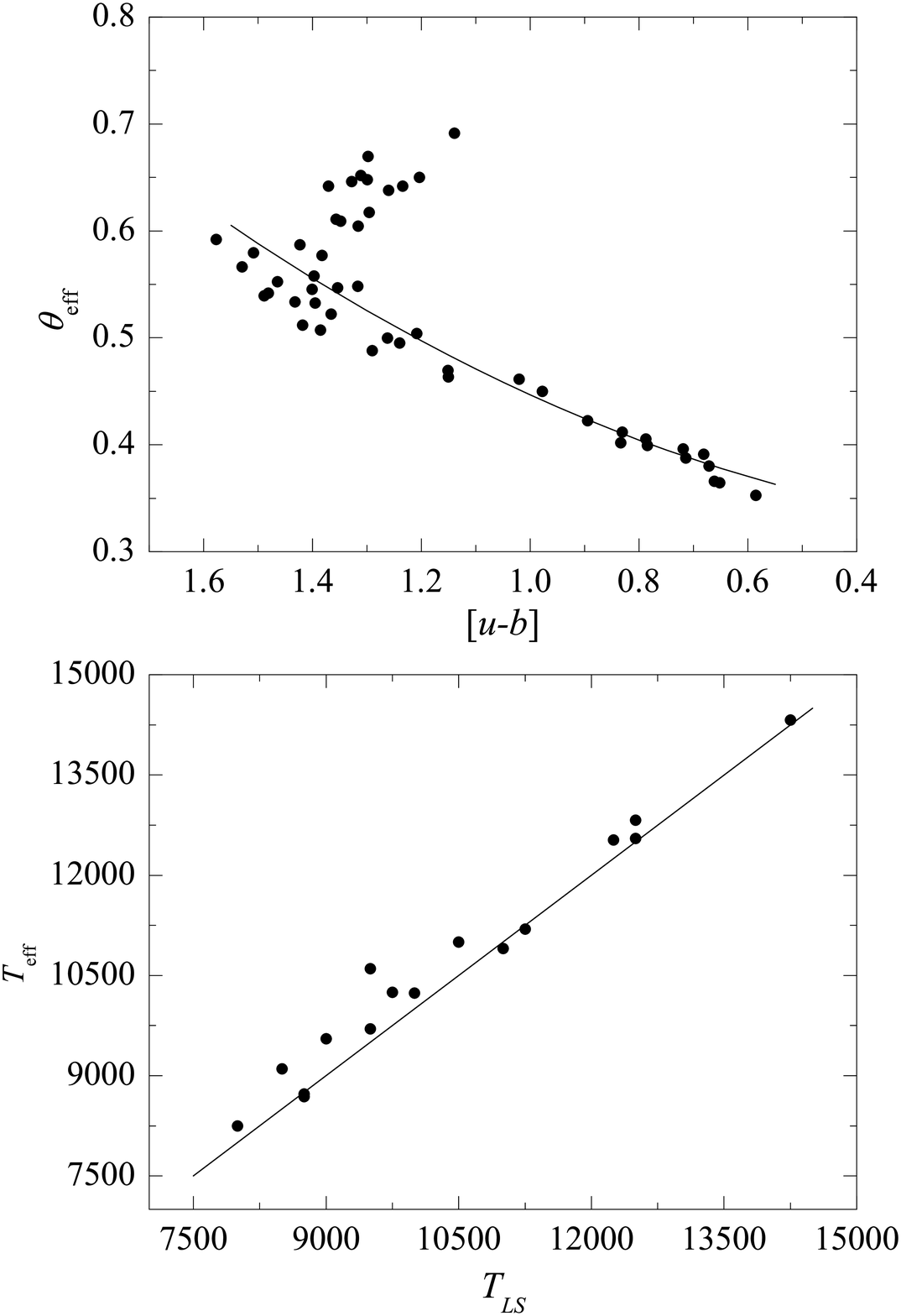}
\caption{The upper panel shows the comparison of the reddening free Str\"omgren index $[u-b]$ to $\theta_\mathrm{eff}$. The line represents the quadratic fit given by \citet{LS08}. the cool stars also are included, although they are not calibratable via the reddening free index. The lower panel shows the discrepancy of the temperatures determined by the authors to our averaged ones without their results. The line gives the one-to-one relation.}
\label{stepien}
\end{center}
\end{figure}

The above defined restricted sample and its $[u-b]$ index is compared to average $\theta_\mathrm{eff}$ values in Fig. \ref{stepien}. We also included the proposed quadratic fit by \citet{LS08}. To clarify the discrepancy, especially for the cooler stars, we compare their determined temperatures to our averaged ones excluding their results, which is shown in the lower panel of Fig. \ref{stepien}. One can see that their temperatures for the cooler stars are somewhat underestimated ($\sim$ 370\,K). Since the differences between the averaged values with or without these results are marginal, we keep their results in our sample; a too rigorous treatment  probably would exclude nearly the entire dataset.

It can be seen in Fig. \ref{stepien} (upper panel) that an uncritical application of relations based on $[u-b]$ or $[c_{1}]$ results in erroneous data for cool stars. The linear fit based on the $[u-b]$ index was therefore limited to stars hotter than about 9000\,K. For this group one can achieve an even more accurate result ($\pm$500\,K) than after applying the correction of $T_{uvby\beta}$. However, without knowledge of additional information like an estimation of temperature via another presented method, it is difficult to separate them. The determined relation is close to former results (\citealt{NA93}, \citealt{S94}), but also to \citet{LS08} who included with their quadratic fit a linear correlation for comparison.  

Temperatures determinable via $Geneva$ $X/Y$ parameters have to be corrected according to Table \ref{relations}, whereas the cooler CP2s can be calibrated using the relation $$\theta_\mathrm{eff}=0.835+0.458(B2-G)_0$$ for normal stars by \citet{HN82} with good agreement. With these two corrections one can achieve an accuracy better than 500\,K for $\sim$85\% of the sample.  

The correction found for $T_{XY}$ is closer to the ``original'' one given by \citet{HN93} than to the revised relation by \citet{HK96}, based on the new grids for $Geneva$ photometry (\citealt{KNK97}) that are also used in the present investigation. The latter one deviates from ours by $\sim \pm$ 400\,K, whereas the ``original'' one differs by $\sim \pm$ 150\,K only at the hot and cool end, respectively.

Stars expected to be hotter than about 9000\,K can be calibrated based on the $Q$-parameter without the need for reddening information. The found $Q$ dependency is in excellent agreement with that given by \citet{M88} based on 11 objects. For cooler stars or if no $(U-B)$ colour information is available, one can also use two relations based on $(B-V)_0$ (Table \ref{relations}). We noticed missing $UB\it{V}$ photometry for numerous CP2 objects. For 11 stars we have not found such data, for 5 stars there is only a lack of $(U-B)$.
 
All photometric systems are capable of estimating effective temperatures of CP2 stars at the same accuracy level with no dependency on temperature. Although it is not possible to separate ``strange'' peculiar objects (like HD~133880) by averaging the results of all possible calibrations, we propose such a method also for the CP2 stars to reduce errors due to photometric uncertainty. 

\begin{figure}[t]
\begin{center}
\includegraphics[width=80mm]{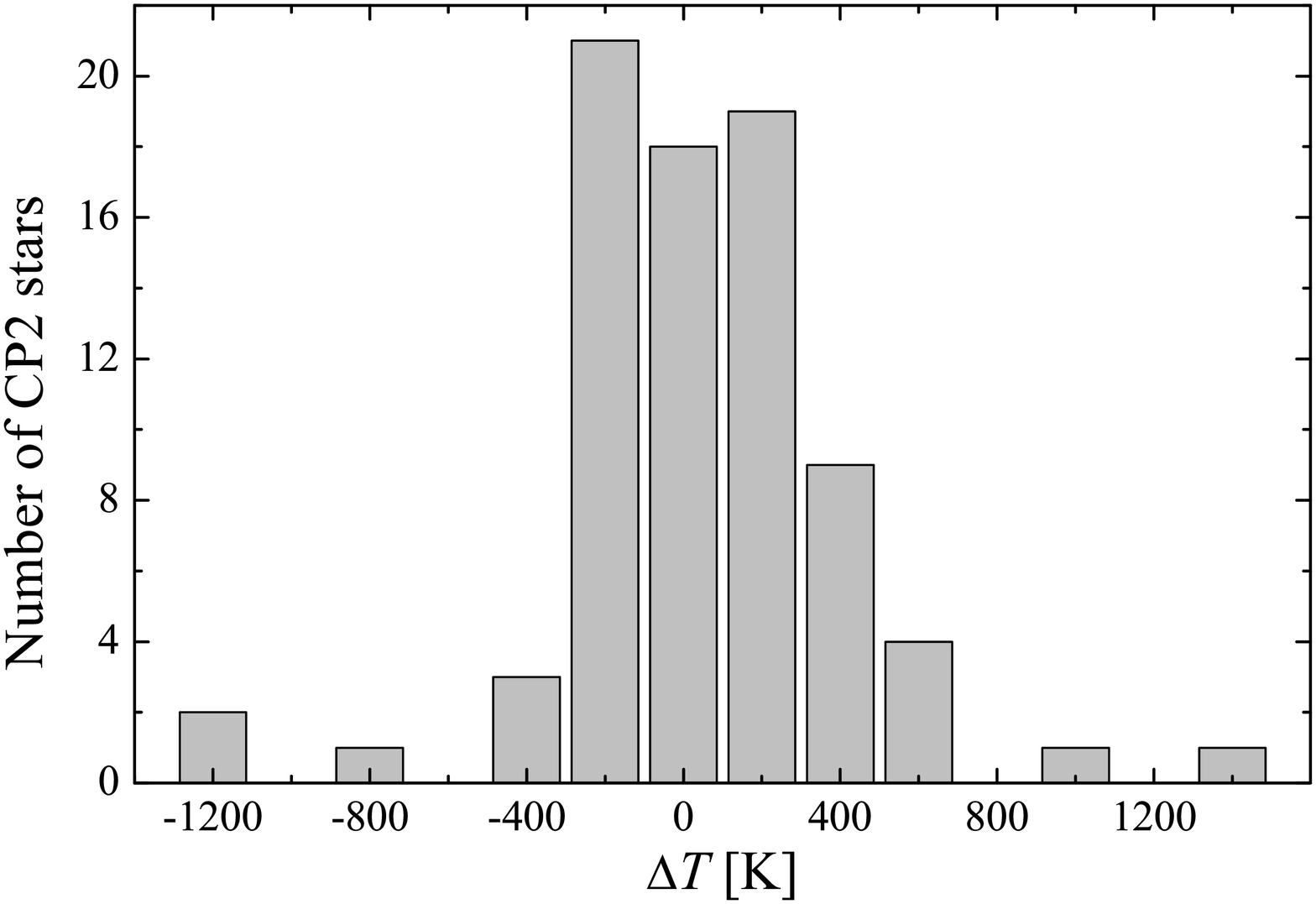}
\caption{Histogram of the temperature deviations for CP2 stars ($T_\mathrm{eff} - \overline{T}_{phot}$).}
\label{CP2temp}
\end{center}
\end{figure}

The $(B-V)_0$ calibration for the hotter stars should be only used as last resort, since the sum of errors of photometry and reddening are not negligible. For intermediate stars ($\sim$11000\,K) 0.02\,mag in total already  results in a temperature difference of $\sim$600\,K. The obtained deviations are presented in Fig. \ref{CP2temp} neglecting the mentioned calibration. We are able to calibrate nearly 90\% of the whole sample within an error of 500\,K, still 75\% better than $\sim$300\,K. 

Among the strongest deviating stars (see Table \ref{deviators}), the abovementioned object was found, as well as HD~215441 (Babcock's star) or HD~157751 for which \citet{HN07} found resolved magnetically split lines and a mean field modulus of 6.6\,kG. The photometrically determined temperatures for this object deviate strongly, only the corrected $T_{uvby\beta}$ result agrees well, whereas the $[u-b]$ relation supplies a temperature already 1750\,K too low. The other star (HD~92499) studied by the authors showing the same effect and even a larger magnetic field modulus is unfortunately among the rejected stars due to the poor photometry available. 

To have an additional comparison for a possible magnetic effect on photometric temperatures, the relationship was deduced by means of the $(B2-G)$ relation above. Since this object is located at a distance of $\sim$280\,pc, an assumption of $E(B-V)=0.05$\,mag is probably the upper limit that leads to a temperature deviating only by 200\,K. For HD~215441, exhibiting the strongest magnetic field, only one direct temperature determination is available to our knowledge (\citealt{LS08}, who discussed the problems of its investigation). 

\subsection{CP3 and CP4}
\label{cp4stars}
Initially analysed separately, no differences between He-weak, He-rich, and CP3 objects were noticed in the gradient of $T_{uvby\beta}$ and $T_{XY}$ results. Therefore these temperature corrections are valid for all CP3 and CP4 members. For the analysis the sample was reduced to stars with at least two temperature determinations, resulting in 14 CP3, 15 CP4a, 9 CP4b, and the two intermediate CP4 objects (see Sect. \ref{collection}). 

HD~137509 exhibits the second largest magnetic field beside Babcock's star (e.g. \citealt{MH97}; \citealt{K06}) and was found to be Si Cr Fe enhanced, but also strongly He underabundant. We therefore included it as a first attempt in the group of He-weak objects. However, like HD~133880 in the CP2 sample, this star was excluded from the analysis because of the large deviation. An inclusion in the CP2 group also would lead to such a rejection. 
The investigation of HD~137509 by \citet{SKK08} shows that appropriate models should be used for the analysis of stars with such a large magnetic field ($\left\langle B \right\rangle \sim 29\,\mathrm{kG}$). They suggest a temperature about 1000\,K higher for this star than previously published, correcting the temperature in the right way according to its deviation, placing it within the scatter of the other representatives in our sample.

Although for the He-rich object HD~60344 two independent temperature determinations are available, it was also excluded from the analysis due to the strongly differing results producing a standard deviation of more than 2000 (see Table \ref{deviators}). The same holds for the CP3 star HD~23408, exhibiting an error of more than 8\%.

\begin{figure}[t]
\begin{center}
\includegraphics[width=80mm]{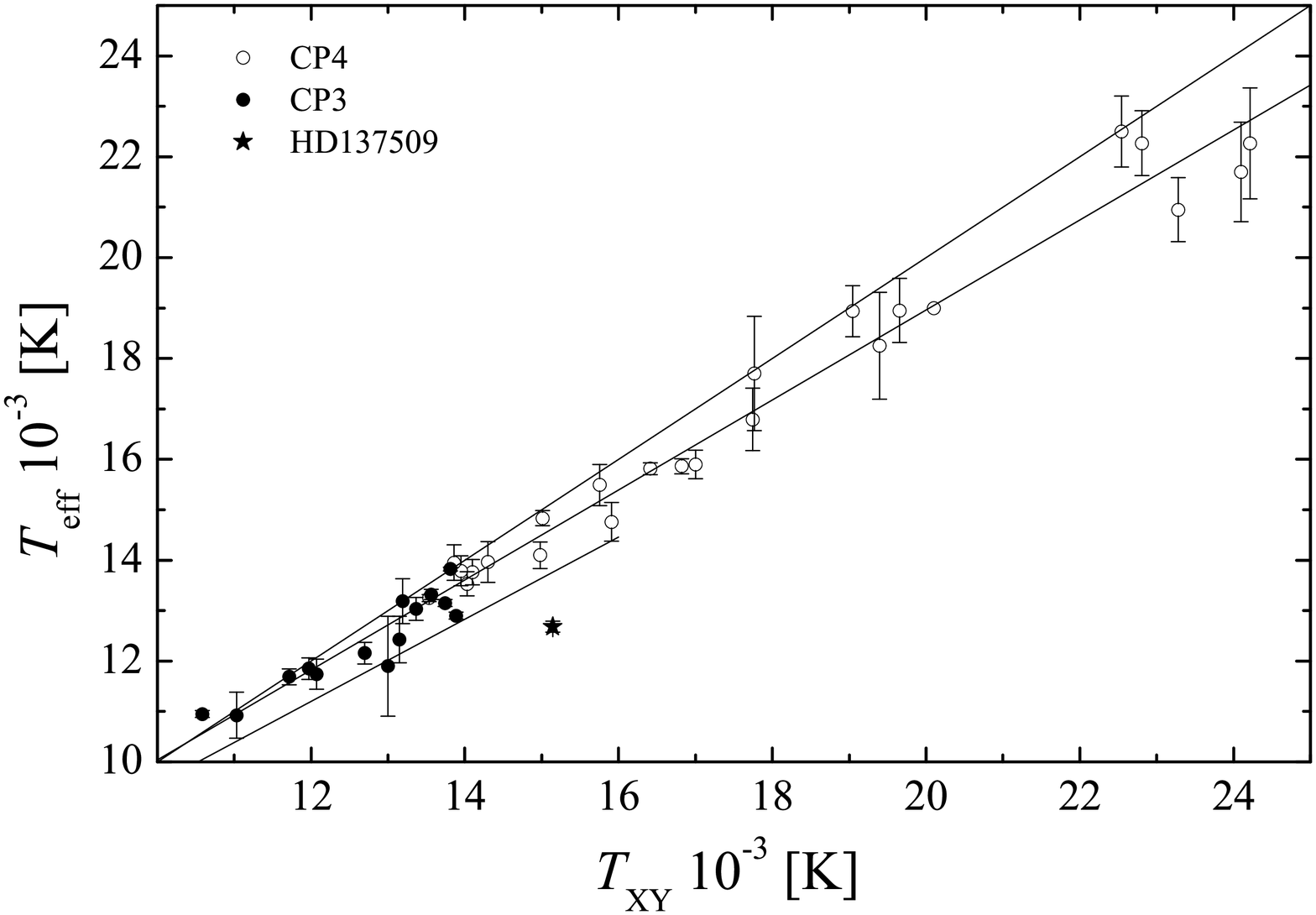}
\caption{Comparison of temperatures determined via $Geneva$ photometry to literature values for the restricted CP3 and CP4 samples. The upper line shows the one-to-one relation, the middle one our linear fit, and the bottom line the CP2 relation for comparison. Additionally, the outstanding object HD~137509 is given. The error bars correspond to the standard deviation of the average effective temperatures.}
\label{CP4genf}
\end{center}
\end{figure}

Temperatures for CP4 objects obtained via the UVBYBETA calibration and $Geneva$ $X/Y$ parameters are found to be overestimated compared to effective temperatures reported in the literature (see Fig. \ref{CP4genf}). Even for CP3 stars a correction of up to 400\,K is necessary, continuing the trend of CP4 stars. A tendency to change with temperature for CP3 and $T_{uvby\beta}$ results was already noticed by \citet{AR00}, and their correction is close to ours. 

Former investigations (e.g. \citealt{HN93}) suggested the direct use of $Geneva$ $X/Y$ temperature for He-weak and HgMn stars, whereas \citet{ZN97} noticed an overestimation of $Geneva$ temperature for He-rich objects. Their proposed correction is in fair agreement with our result. However, in contrast to our study they included similar errors for both axes to determine the relation, but did not list the errors used. In Fig. \ref{CP4genf} one can see that a direct use of the $Geneva$ results significantly overestimates the temperature.  

Due to the limited number of available cluster CP3s, a comparison to the photometric colour excess estimations is not significant. Since there are only a few objects closer than 100\,pc, a temperature calibration based on the Johnson $(B-V)$ index, very probably affected by interstellar reddening, was not undertaken. For the CP4 group also only the reddening free $Q$-parameter was investigated, as at such high temperatures even a small reddening error already results in a large deviation (see also Sect. \ref{lipski}).

Inspecting the reddening free $Q$ and $[u-b]$ indices, the different properties of CP3 and CP4 stars are apparent. Among the CP3 objects two sequences are noticed, one following the He representatives. Therefore the CP3 sample was divided into two groups, the ``classical'' cooler HgMn stars (CP3a) and predominantly hotter ones (CP3b). Members of the latter group can be mostly considered as PGa objects, the hotter extension of HgMn stars, exhibiting deficient He and strongly overabundant P and Ga (\citealt{HG07}; \citealt{RLR06}). We have noticed that several of these objects were studied as pure He-weak objects (e.g. by \citealt{CAT07}). Additional publications helped to clarify their membership to the CP3 (sub)group. However, since the temperature behaviour of CP3b and CP4 members is similar in all investigated systems, an ambiguous classification fortunately does not influence the resulting temperature.

Using the reduced samples mentioned above, the CP4 are combined with the three members of the CP3b group to determine the relations for $Q$ and $[u-b]$. The exclusion of the latter group does not alter the results listed in Table \ref{relations}. Due to the limited number of cooler HgMn objects with more than one temperature determination, the whole sample of 18 stars was used for the analysis to better cover the temperature range.

The $UB\it{V}$ and $[u-b]$ calibrations applied to HD~137509 agree well with the temperature of 13750\,K proposed by \citet{SKK08}. However, that can be also by chance due to its variability. An amplitude of about 0.1\,mag was found in the $Geneva$ [U] colour by \citet{ML97}. All other results indicate a consistently higher temperature for this star after the corrections (14630 and 14500\,K for $Geneva$ and $uvby\beta$ photometry, respectively). See also Sect. \ref{misclass} in this respect.

For the hot CP4b star CPD$-$62 2124, we noticed large differences between the Str\"omgren and $UB\it{V}$ results. Since this star agrees very well with the Johnson $Q$ relation, the discrepancy is probably caused by the Str\"omgren photometry taken from \citet{PL86}, therefore this result is rejected. Unfortunately, no $Geneva$ photometry is available for a comparison in this system. However, \citet{ZN97} noted, that this star shows emission at least in its Balmer lines and the helium abundance is less reliable. Additionally, the emission was not only seen in the star but also in the surrounding sky, caused by a nebula.

\begin{figure}[t]
\begin{center}
\includegraphics[width=80mm]{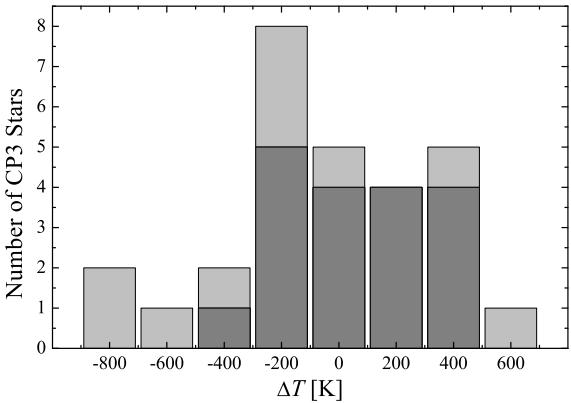}
\caption{Histogram of the temperature deviations for CP3 stars ($T_\mathrm{eff} - \overline{T}_{phot}$). The dark grey portion represents the ``classical'' HgMn objects.}
\label{cp3temp}
\end{center}
\end{figure}

\begin{figure}[t]
\begin{center}
\includegraphics[width=80mm]{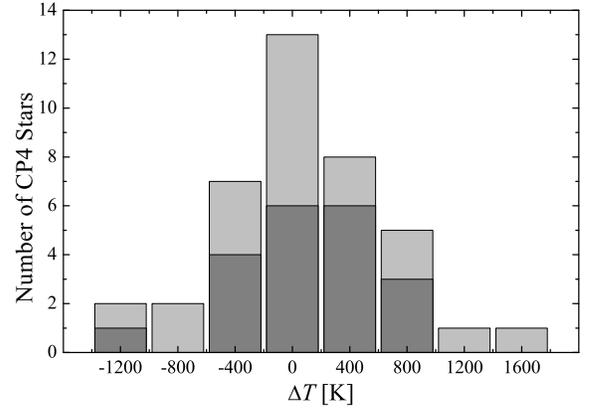}
\caption{Histogram of the temperature deviations for CP4 stars ($T_\mathrm{eff} - \overline{T}_{phot}$). The dark grey portion represents the He-weak objects.}
\label{cp4temp}
\end{center}
\end{figure}

The deviations of the determined temperatures compared to literature values are presented in Figs. \ref{cp3temp} and \ref{cp4temp} for CP3 and CP4 stars respectively. 
All investigated photometric systems show the same ability to calibrate effective temperatures for the individual subgroups.
An average of all results does not provide an improvement in accuracy, but helps to avoid erroneous temperatures due to wrong photometry in a single system. About 85\% of the CP3 stars are calibrated to better than 500\,K, for the subgroup of HgMn objects even a slightly better result ($\sim 400$\,K) can be obtained. In the case of CP4 members, one can achieve an accuracy of $\sim$700\,K. However, in all groups some stronger deviating objects are found, which are discussed in Sect. \ref{misclass}.

Like the Am group, CP3 stars are often part of spectroscopic binaries. \citet{AR00}, a main contributor to our sample, overcame this problem by studying single objects or those whose companions have not been detected or contribute very little to the optical region fluxes. A restriction to objects studied in this reference (10 stars) reduces the error range to $\sim 300$\,K. However, this CP group still suffers from a lack of well investigated stars.   

In Tables \ref{cp3list} and \ref{cp4list} the compiled data are listed. For some objects, the peculiarity type given by \citet{R91} is misleading. The following references were therefore used for the classification: HD~35497 and HD~77350 (\citealt{AR00}), HD~147550 (\citealt{LA94}), HD~19400 (\citealt{HNS06}), HD~144667 (\citealt{CH07}).
 
\begin{table*}
\begin{minipage}[t]{\textwidth}

\caption{Overview of the temperature calibrations for the individual CP groups and photometric systems determined in this study.} 
\label{relations} 
\centering 
\tiny
\renewcommand{\footnoterule}{}
\begin{tabular}{l l l l l l l} 
\hline\hline 
CP Type & System & Relation & Errors\footnote{Errors of the linear fits} & Restriction & $R$\footnote{Correlation Coefficient} & \# Stars\\ 
\hline 
CP1 & $UB\it{V}$    & $\theta_\mathrm{eff} = 0.527 + 0.515(B-V)_{0}$ & 0.003/0.013 &   & 0.993 & 24\\
    & $Geneva$ & $\theta_\mathrm{eff} = 0.632 + 0.640(B2-V1)_{0}\footnote{taken from \cite{HAU85}}$ & & $-0.160 \leq (B2-V1)_{0} \leq +0.730$ & &\\
    &          & direct use of $T_{XY}$  & & $(B2-V1)_{0} < -0.160$ & &\\    
	& $uvby\beta$  & direct use of $T_{uvby\beta}$ & & & &\\
CP2 & $UB\it{V}$    & $\theta_\mathrm{eff} = 0.541 + 0.389Q$ & 0.004/0.014 & $T_\mathrm{eff} \ga 9000$  & 0.982 & 29\\
    &          & $\theta_\mathrm{eff} = 0.572 + 1.177(B-V)_{0}$ & 0.011/0.089 & $-0.20 \leq (B-V)_{0} \leq -0.05$  & 0.948 & 21 \\
    &          & $\theta_\mathrm{eff} = 0.542 + 0.388(B-V)_{0}$ & 0.005/0.030 & $-0.05 \leq (B-V)_{0} \leq 0.40$  & 0.932 & 28 \\
    & $Geneva$ & $\theta_\mathrm{eff} = 0.835 + 0.458(B2-G)_{0}$\footnote{taken from \cite{HN82}} & & $T_\mathrm{eff} \la 9000$ & &\\
    &          & $T_\mathrm{eff}=1420 + 0.815T_{XY}$  & 280/0.023 & $T_\mathrm{eff} \ga 9000$ & 0.988 & 30\\
    & $uvby\beta$  & direct use of $T_{uvby\beta}$  &  & $T_{uvby\beta} < 9000 $ & & \\
    &   & $T_\mathrm{eff}=2090+0.756T_{uvby\beta}$ & 300/0.025 & $T_{uvby\beta} \geq 9000 $ & 0.984 & 31\\
    &              & $\theta_\mathrm{eff}=0.234+0.213[u-b]$ & 0.009/0.008 & $T_\mathrm{eff} \ga 9000$ & 0.980 & 33\\
CP3/4& $Geneva$ & $T_\mathrm{eff}=1120 + 0.892T_{XY}$ & 350/0.021 & & 0.990 & 37\\
CP3/4& $uvby\beta$ & $T_\mathrm{eff}=2230+0.809T_{uvby\beta}$ & 300/0.018 & & 0.991 & 37\\
CP3a  & $UB\it{V}$ & $\theta_\mathrm{eff} = 0.501 + 0.323Q$ & 0.007/0.026 & & 0.953 & 18\\
     &  $uvby\beta$  & $\theta_\mathrm{eff}=0.233+0.196[u-b]$ & 0.014/0.014 & & 0.965 & 17\\
CP3b/4  & $UB\it{V}$ & $\theta_\mathrm{eff} = 0.540 + 0.418Q$ & 0.009/0.017 & & 0.980 & 27\\
     &  $uvby\beta$  & $\theta_\mathrm{eff}=0.173+0.286[u-b]$ & 0.005/0.008 & & 0.989 & 27 \\
\hline 
\end{tabular}
\end{minipage}
\end{table*}

\subsection{Strong deviating objects}
\label{misclass}
In Table \ref{deviators} one can find the strongest deviating stars after applying individual corrections.  
For the star HD~37470 the standard temperature calibrations result in a much better agreement than the corrected values. The same holds for DM $-$27 3748 and HD~21699. However, the latter object is a striking helium and silicon variable (\citealt{HM94}) and the deviation of the other is still below 5\%. For HD~51688, showing enhanced Si Mn, but also a He-weak behaviour, the CP2 corrections would reduce the deviation significantly to $-290$\,K. The object HD~66522 also agrees much better as a CP2, although its high temperature excludes it from this group. In Sect. \ref{cp4stars} large differences between the individual photometric results for the strong magnetic object HD~137509 were noticed, which are not found to such an extent using the CP2 relations. Also the deviation from literature values are noticeably reduced to $-720$\,K, when treating it as CP2. If using the study by \citet{SKK08} for comparison, the temperature difference decreases to +360\,K, but the use of the CP4 corrections also results in a good agreement with the reference above. However, the more homogeneous photometric results suggest that this object should be classified as CP2. It seems that strong magnetic fields do not have such a large influence on the photometric temperatures. For the other strong magnetic CP2 stars HD~215441, HD~133880, and HD~157751, an investigation like in \citet{SKK08} would be useful to obtain more comparative values. No influence on the temperature calibrations is given due to probable misclassifications, since none of these objects were used for the calculations. Nevertheless, a spectroscopic re-investigation of the listed stars is necessary, in order to clarify their status and temperature. For most of them only one temperature determination or strongly differing result is found in the literature.

\begin{table}
\caption{The strongest deviating stars after applying the individual corrections.} 
\label{deviators} 
\centering 
\begin{tabular}{r l l l } 
\hline\hline 
HD/DM& CP & $T_\mathrm{eff} / \sigma$ & $\Delta T/\sigma \overline{T}_\mathrm{phot}$  \\ 
\hline 
$-$27 3748 & CP4b & 23000 & 1120/280 \\
21699 & CP4a & 16000 & 950/100 \\
23408 & CP3b & 11900/990 & $-$800/90 \\
26571 & CP2 & 11750 & $-$1170/200 \\
37470 & CP2 & 13000 & 1340/170 \\
51688 & CP3b & 12500 & $-$880/80 \\
60344 & CP4b & 22500/2120 & 1490/260 \\
66522 & CP4b & 18000 & $-$1210/90 \\
133880 & CP2 & 10700/60 & $-$1230/210 \\
137509 & CP4a & 12680/110 & $-$1350/680 \\
157751 & CP2  & 11300 & 1040/790 \\
215441 & CP2  & 14000 & $-$780/390 \\
\hline 
\end{tabular}
\end{table} 

\section{Bolometric correction}

Two studies (\citealt{LB07}; \citealt{LS08}) examined the bolometric correction ($BC$) for magnetic CP stars. Both investigations presented a relation based on effective temperature, a more appropriate solution than the one by \citet{L84} on the basis of the $Geneva$ $(B2-G)$ colour index. While the first reference shows a comparison of published $BC$s to photometrically determined temperatures, the second authors used their own results for the correlation, whose temperature determinations were discussed in Sect. \ref{lipski}. Since we have shown that the former temperature corrections are partially inaccurate, we checked the validity of the two findings using our sample of averaged effective temperatures. For this purpose we collected the integrated fluxes ($F_{t}$) for objects in our sample reported in the references above, and some more by \citet{SB79}, \citet{SB85}, \citet{GLU87}, \citet{MM92}, \citet{CAT07}, and \citet{BNC08} to build averages whenever possible and calculated a mean $BC$ using $$BC = -2.5\,\log(F_{t})-m_{v}-11.49.$$ In total, we found 85 individual fluxes for 35 CP2, 7 CP3 and 11 CP4 objects. To be as homogeneous as possible, the visual magnitude ($m_{v}$) of the $Geneva$ photometric system was used whenever available, corrected for interstellar absorption by using the determined reddening values and a mean total-to-selective ratio of 3.1. 

For HD~22920 we noticed that the integrated flux listed by \citet{L84} is most likely an error, the value does not correspond to the given bolometric correction. We therefore used the tabulated $BC$ value directly for this star, since it perfectly agrees with the compiled temperature.

\citet{LB07} argued that the integrated fluxes by \citet{SD89} are not corrected for reddening and removed the hottest (farthermost) stars from their sample. We therefore checked all objects by means of the determined reddening values. All objects in references with no explicitly given reddening values are not significantly reddened ($< 0.03$\,mag), but we rejected four deviating (CP2) objects by comparing the listed $E(B-V)$ values in the work by \citet{LS08} to ours. For the remaining 42 magnetic CP objects the second order fit $$BC = -5.737+18.685\theta_\mathrm{eff}-15.135\theta_\mathrm{eff}^2,$$ valid for 7500-19000\,K, best represents their behaviour. Even the restriction to nearly non reddened ($< 0.03$\,mag) stars or the use of objects with at least two integrated flux measurements does not alter the result significantly, but the cool part especially  is not sufficiently covered by such data.

\begin{figure}[t]
\begin{center}
\includegraphics[width=90mm]{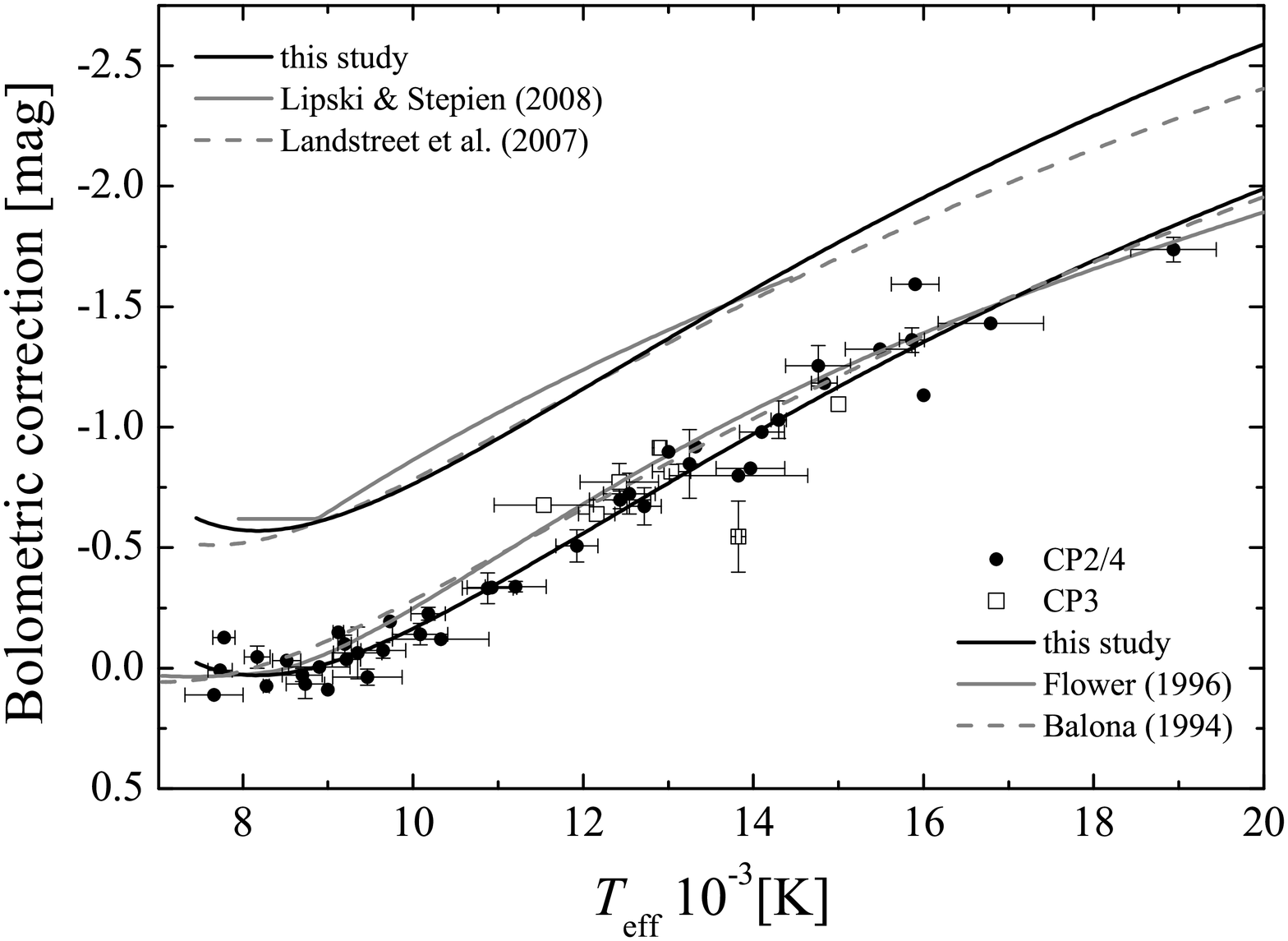}
\caption{Bolometric corrections for CP stars together with our fit (solid black line) and the relations for normal stars by \citet{BAL94} and \citet{FLO96}. For better presentation, the comparison of our fit to \citet{LS08} and \citet{LB07} is shifted by $-0.6$\,mag. Whenever available, the error of BC according to error propagation using the standard deviation of the averaged integrated fluxes and 0.02\,mag in reddening is given. Additionally the standard deviation of the averaged temperatures is shown.}
\label{bol}
\end{center}
\end{figure}

The data are presented in Fig. \ref{bol} together with the relation determined above. No difference was noticed between magnetic CP2 and CP4 objects, however only a small temperature overlap between the two groups is available. Within the temperature range of 7500-14000\,K an uncertainty of 0.1\,mag applies; for hotter stars up to $\sim$19000\,K a slightly higher value of 0.15\,mag has to be taken into account. The discrepancy of $-$0.16\,mag with the result by \citet{LB07} at the hot end (Fig. \ref{bol}) is probably caused by the use of overestimated photometric temperatures for He-weak stars in this reference (see Sect. \ref{cp4stars} and Fig. \ref{CP4genf} in this respect) that lowers the resulting slope. However, the large scatter and low number of He objects still prevents a clear conclusion for such hot CP representatives, but at least the main CP2 temperature domain up to 14000\,K seems to be well defined. In contrast to the reference above, the relation by \citet{LS08} is shifted to larger negative $BC$ values (see Fig. \ref{bol}) in this temperature region, placing it close to the $BC$ of normal stars. This can be explained due to the lower temperatures mentioned in Sect. \ref{lipski}, by the use of a zero reddening for all stars closer than 100\,pc and $E(B-V) \le 0.03$\,mag as well as by the difference of the constant used to transform integrated fluxes to the bolometric correction. The bolometric correction for normal stars by \citet{BAL94} and \citet{FLO96} is given as a comparison in Fig. \ref{bol}. The difference in $BC$ between normal and CP stars of same temperature is not larger than about 0.1\,mag, in agreement with \citet{KS08} who determined using model fluxes a systematic difference of the same value.

Concerning CP3 stars, the situation is even more problematic because of the extremely limited number. For the two closest stars ($\la 50$\,pc) zero reddening was assumed, for the others the values by \citet{L84} and \citet{CAT07} were adopted. One strongly deviating object (HD 358) can be found in Fig. \ref{bol}, the others are placed around the normal star relation, but we conclude that the available data has to be increased significantly, which also holds for the magnetic CP types, especially for the hotter representatives.   

\section{Conclusions}

The literature was consulted to compile a sample of 176 chemically peculiar stars with temperature determinations other than from photometric results, in order to calibrate photometrically determined temperatures. The obtained corrections and relations listed in Table \ref{relations} are therefore based on a much larger sample than in previous studies. A clear CP classification is necessary in order to use the proper correction and to obtain the best possible result. Except the $(B-V)_{0}$ relation for hot CP2 stars (see Table \ref{relations}), all presented corrections within the respective CP subgroups are able to determine the effective temperature at about the same accuracy level. However, to avoid erroneous results due to incorrect photometry in a single system, a mean value of all available individual results should be always used. 

A comparison to former results showed that He-weak and He-rich objects especially have been overestimated until now, influencing all previous studies (e.g. \citealt{LB07}). The new lower temperatures and bolometric corrections will place their hotter (CP4) objects at somewhat older ages, on the border of the adopted errors. These authors also discussed the problems of photometric temperatures (see Sect. 3.2.2. in the reference) referring to the work by \citet{KS06} who conclude that photometric temperature calibrations for normal stars are not far from the true values of CP stars. However, the present study confirms that compared to normal stars, sometimes considerable corrections are necessary (except cool Am and CP2 stars), although we cannot exclude effects influencing all compiled temperatures determined via different methods. For example, the effect of strong magnetic fields has to be examined on a much larger sample. On the other hand, the presented results for HD~137509 together with the work by \citet{SKK08} would seem to suggest that photometric temperatures are not significantly affected.

Recently, \citet{BNC08} obtained the first real direct temperature measurement for the roAp HD~128898 by means of angular diameter and integrated flux measurements. Their investigation yields a 480\,K lower temperature compared to the spectral analysis by \citet{KRW96}. For a final solution of the problematic CP temperature calibration, more studies of this kind are needed.

It was shown that photometric standard calibrations can be applied to determine interstellar reddening also for CP2 and CP4 objects. 
Additionally, a revised bolometric correction for magnetic CP stars is presented. Together with the proposed photometric temperature relations it can serve to study the evolutionary status of these objects to the highest possible accuracy.

The Tables \ref{cp1list}$-$\ref{cp4list}, available in electronic form at A\&A, list the compiled objects of the individual CP groups together with average temperature (literature value $\overline{T}_\mathrm{eff}$ and photometrically determined $\overline{T}_\mathrm{phot}$), number of literature sources \#, the deviation $\Delta T$, visual magnitude $m_{v}$, the revised Hipparcos parallaxes $\pi$, and the bolometric corrections found in the literature. The existing photometry is given in the form $uvby\beta/Geneva/UB\it{V}$. For the magnetic subgroups the calculated reddening also is shown. In parentheses the errors of the last significant digits are given. The peculiarity type listed in the CP2-4 sample is taken from \citet{R91}. Table \ref{cpref} lists the references found for the individual stars, as well as the method used for the temperature determination.

\onltab{4}{
\begin{table*}
\begin{minipage}[t]{180mm}
\caption{The CP1 (Am) stars.} 
\label{cp1list} 
\centering 
\scriptsize{
\begin{tabular}{r r c r l c l c r} 
\hline\hline 
HD & HIP & $m_{v}$  & $\pi$  & $\overline{T}_\mathrm{phot}$  & Phot. & $\overline{T}_\mathrm{eff}$  & \# & $\Delta T$ \\
&  &  [mag] &  [mas] &  [K] &  &  [K] &  &  [K] \\
\hline 
27628&20400&5.71&22.53(55)&7260(50)&+/+/+&7210(90)&2&$-$50 \\
27749&20484&5.63&20.04(48)&7410(30)&+/+/+&7340(180)&5&$-$70 \\
27962&20648&4.29&21.96(51)&9040(110)&+/+/+&9070(210)&3&30 \\
28226&20842&5.71&21.25(41)&7470(100)&+/+/+&7450&1&$-$20 \\
28355&20901&5.02&20.47(28)&7880(70)&+/+/+&7950&1&70 \\
28527&21029&4.78&23.15(31)&8110(70)&+/+/+&8000(100)&2&$-$110 \\
28546&21039&5.48&22.27(49)&7640(20)&+/+/+&7640(140)&2&0 \\
29140&21402&4.24&20.86(94)&8010(170)&+/+/+&7940&1&$-$70 \\
29499&21670&5.38&20.33(36)&7650(10)&+/+/+&7690&1&40 \\
30210&22157&5.35&13.85(38)&8040(120)&+/+/+&8100&1&60 \\
33204&24019&5.93&19.04(134)&7490(110)&+/+/+&7650&1&160 \\
33254&23983&5.42&18.27(36)&7740(40)&+/+/+&7760(190)&2&20 \\
58142&36145&4.60&11.92(24)&9530(170)&+/+/+&9500&1&$-$30 \\
67523&39757&2.82&51.34(15)&6820(70)&+/+/+&6700&1&$-$120 \\
78362&45075&4.65&25.82(54)&7170(100)&+/+/+&7220(230)&2&50 \\
94334&53295&4.67&13.22(50)&9950(80)&+/+/+&10030&1&80 \\
95418&53910&2.36&40.89(16)&9560(170)&+/+/+&9600(10)&2&40 \\
95608&53954&4.42&25.67(17)&9040(70)&+/+/+&8950(430)&2&$-$90 \\
97633&54879&3.32&19.77(17)&9400(240)&+/+/+&9250(180)&2&$-$150 \\
141795&77622&3.70&46.28(19)&8360(20)&+/+/+&8420&1&60 \\
162132&87045&6.47&8.56(40)&8770(50)&+/+/+&8800&1&30 \\
173648&91971&4.33&20.88(17)&8040(30)&+/+/+&8160&1&120 \\
182564&95081&4.58&14.24(12)&9210(140)&$-$/+/+&9130&1&$-$80\\
188728&98103&5.27&14.85(24)&9450(100)&+/+/+&9530&1&80 \\
189849&98543&4.66&13.79(20)&8000(120)&+/+/+&7820(110)&3&$-$180 \\
196724&101867&4.80&15.56(52)&10130(280)&+/+/+&10200&1&70 \\
206088&106985&3.67&20.76(72)&7330(60)&+/+/+&7640&1&310 \\
209625&108991&5.28&14.68(30)&7780(70)&+/+/+&7820(100)&3&40 \\
213320&111123&4.81&11.27(125)&10180(50)&+/+/+&10130&1&$-$50 \\
214994&112051&4.80&10.93(67)&9580(120)&+/+/+&9590(20)&3&10 \\

\hline 
\end{tabular}
}
\end{minipage}
\end{table*} 
}

\onltab{5}{
\begin{table*}
\begin{minipage}[t]{180mm}
\caption{The CP2 stars.} 
\label{cp2list} 
\centering 
\renewcommand{\footnoterule}{}
\scriptsize{
\begin{tabular}{r r l c r l l c l c r l } 
\hline\hline 
HD&HIP&Pec. Type& $m_{v}$ &$\pi$ &$E(B-V)$ &$\overline{T}_\mathrm{phot}$ &Phot.&$\overline{T}_\mathrm{eff}$ &\#& $\Delta T$ & $BC$ \\
&&&[mag]&[mas]&[mag]& [K]&&[K]&& [K] & [mag] \\
\hline 
8441&6560&Sr&6.69&4.88(59)&0&9140(190)&+/+/+&9200&1&60 &  \\
9484&7222&Si&6.58&8.54(43)&0(1)&9980(160)&+/+/+&10200&1&220 &   \\
12098&&\footnote{roAp according to \citet{RNW04}.}&7.97&&0&7670&+/$-$/$-$&7800&1&130 &   \\
12767&9677&Si&4.68&8.79(26)&0&12860(160)&+/+/+&13000(340)&3&140 &  \\
15089&11569&Sr&4.47&24.55(81)&0&8390(200)&+/+/+&8280(40)&2&$-$110 & 0.07   \\
18610&13534&Cr Eu Sr&8.15&4.68(54)&0&7680(340)&+/+/$-$&8100&1&420 &   \\
19832&14893&Si&5.78&6.47(76)&0.02(1)&12440(120)&+/+/+&12430(360)&3&$-$10 & $-$0.70(4)   \\
23387&&Cr Si&7.18&&0.05&8420(230)&+/+/+&8250&1&$-$170 &  \\
24155&18033&Si&6.30&8.13(35)&0.09(1)&13450(100)&+/+/+&13780(70)&2&330 & \\
24712&18339&Sr Eu Cr&5.99&20.32(39)&0&7300(220)&+/+/+&7290(60)&2&$-$10 &  \\
25823&19171&Sr Si&5.19&7.76(36)&0.01(1)&12890(90)&+/+/+&12720(200)&3&$-$170 & $-$0.67(8)  \\
26571&19672&Si&6.13&3.76(44)&0.34(1)&12920(200)&+/+/+&11750&1&$-$1170 &   \\
27309&20186&Si Cr&5.37&9.99(29)&0&12000(200)&+/+/+&11930(250)&2&$-$70 & $-$0.51(7) \\
32549&23607&Si Cr&4.67&8.95(24)&0(1)&9850(150)&+/+/+&9730&1&$-$120 & $-$0.19  \\
34452&24799&Si&5.38&7.91(36)&0&13980(340)&+/+/+&13830(810)&2&$-$150 & $-$0.80  \\
37470&26530&Si&8.22&3.41(88)&0.15(2)&11660(170)&+/+/+&13000&1&1340 & $-$0.90  \\
37808&26728&Si&6.45&4.11(40)&0.01(2)&13140(190)&+/+/+&12890(200)&3&$-$250 &   \\
40312&28380&Si&2.64&19.71(16)&0&10180(40)&+/+/+&10180(200)&4&0 & $-$0.23(3) \\
43819&30019&Si&6.27&3.85(79)&0.01(1)&11240(160)&+/+/+&10930(290)&4&$-$310 & $-$0.34  \\
60435&36537&Sr Eu&8.90&4.40(76)&0.06&8230(140)&+/+/$-$&8100&1&$-$130 &   \\
62140&37934&Sr Eu&6.47&10.35(45)&0.02&7920(180)&+/+/+&7800(140)&2&$-$120 &  \\
65339&39261&Eu Cr&6.03&10.14(52)&0&8250(160)&+/+/+&8170(150)&3&$-$80 & $-$0.05(5)  \\
71866&41782&Eu Sr Si&6.73&7.52(46)&0&9240(470)&+/+/+&9000&1&$-$240 &   \\
75445&43257&Sr Eu&7.14&9.23(45)&0&7560(30)&+/+/$-$&7700&1&140 &   \\
81009&45999&Cr Sr Si&6.52&6.92(60)&0&7970(120)&+/+/+&8250(350)&2&280 &   \\
90569&51213&Sr Cr Si&6.01&7.73(30)&0&9850(150)&+/+/+&10500&1&650 &   \\
92664&52221&Si&5.50&6.23(24)&0(1)&13960(70)&+/+/+&14300(90)&3&340 & $-$1.03(8)  \\
94427&53290&Sr&7.37&6.96(62)&0.03&7300(70)&+/+/$-$&7500&1&200 &  \\
108662&60904&Sr Cr Eu&5.26&13.72(25)&0.01(1)&10050(200)&+/+/+&10330(570)&4&280 & $-$0.12  \\
108945&61071&Sr&5.45&12.09(27)&0&8870(130)&+/+/+&8700(240)&6&$-$170 & 0.03(3)  \\
110066&61748&Sr Cr Eu&6.41&7.43(39)&0&9180(350)&+/+/+&9030(60)&3&$-$150 &  \\
111133&62376&Sr Cr Eu&6.32&3.76(40)&0&9670(280)&+/+/+&9850(220)&2&180 &   \\
112185&62956&Cr Eu Mn&1.75&39.50(20)&0&9190(200)&+/+/+&9350(290)&4&160 & $-$0.06(11)  \\
112413&63125&Eu Si Cr&2.84&28.42(89)&0&11480(200)&+/+/+&11210(360)&5&$-$270 & $-$0.34(2)  \\
115708&64936&Sr Eu&7.79&8.64(79)&0.01&7660(220)&+/+/+&7760(350)&2&100 &   \\
116114&65203&Sr Cr Eu&7.03&7.71(55)&0.07&7800(60)&+/+/$-$&7850(210)&2&50 &  \\
118022&66200&Cr Eu Sr&4.92&17.65(20)&0.02(2)&9420(50)&+/+/+&9460(410)&6&40 & 0.04(3)  \\
120198&67231&Eu Cr&5.67&11.23(23)&0&10090(290)&+/+/+&10080(330)&3&$-$10 & $-$0.14(4)  \\
124224&69389&Si&4.99&12.63(21)&0(1)&12120(110)&+/+/+&12540(300)&6&420 & $-$0.72(8) \\
125248&69929&Eu Cr&5.86&9.81(68)&0.02(2)&9850(230)&+/+/+&9650(260)&4&$-$200 & $-$0.07(3)  \\
126515&70553&Cr Sr&7.09&9.39(62)&0.01(1)&9640(160)&+/+/+&9500&1&$-$140 &   \\
128898&71908&Sr Eu&3.17&60.36(14)&0&7820(130)&+/+/+&7660(340)&2&$-$160 & 0.11  \\
133029&73454&Si Cr Sr&6.36&5.89(28)&0&10750(170)&+/+/+&10880(300)&4&130 & $-$0.33(6)  \\
133792&74181&Sr Cr&6.25&5.50(43)&0&9030(260)&+/+/$-$&9300(140)&2&270 &   \\
133880&74066&Si&5.78&9.03(33)&0&11930(210)&+/+/+&10700(60)&2&$-$1230 &   \\
134305&74109&Sr Eu Cr&7.24&6.64(67)&0.02&8070(170)&+/+/+&8200&1&130 &   \\
137909&75695&Sr Eu Cr&3.67&29.17(76)&0&7710(260)&+/+/+&8340(360)&4&630 &   \\
137949&75848&Sr Eu Cr&6.66&11.27(67)&0&7420(410)&+/+/+&7530(40)&2&110 &   \\
140160&76866&Sr&5.32&14.83(41)&0&9120(110)&+/+/+&9120(60)&3&0 & $-$0.15  \\
144897&79197&Eu Cr&8.60&5.62(103)&0.31(2)&11140(140)&$-$/+/+&11250&1&110 & \\
148112&80463&Cr Eu&4.58&13.04(64)&0.03&9520(100)&+/+/+&9220(160)&2&$-$300 & $-$0.04  \\
149822&81337&Si Cr&6.38&7.92(38)&0&10430(250)&+/+/+&10750&1&320 &  \\
151525&82216&Eu Cr&5.21&8.29(27)&0&9360(130)&+/+/+&9240(130)&2&$-$120 &  \\
152107&82321&Sr Cr Eu&4.80&18.10(34)&0&8760(30)&+/+/+&8730(230)&3&$-$30 & 0.07(6)  \\
153882&83308&Cr Eu&6.28&6.15(44)&0.03(3)&9250(190)&+/+/+&9450(580)&3&200 &  \\
155102&83816&Si&6.35&7.38(44)&0&9140(190)&+/$-$/+&9000&1&$-$140 &   \\
157751&85372&Si Cr&7.65&6.20(74)&0&10260(790)&+/+/$-$&11300&1&1040 &   \\
166473&&Sr Eu Sr&7.94&&0.04&7500(470)&+/+/$-$&7850(210)&2&350 &  \\
168733&90074&Ti Sr&5.33&5.84(33)&0.03(1)&12730(210)&+/+/$-$&13320&1&590 & $-$0.92  \\
170973&90858&Si Cr Sr&6.42&3.29(40)&0.05(2)&10830(230)&+/+/+&10740(20)&2&$-$90 &  \\
171247&90971&Si&6.41&1.77(41)&0.08(1)&11630(340)&+/+/+&12170&1&540 &   \\
171782&91224&Si Cr Eu&7.84&2.00(84)&0.12(2)&11310(90)&+/$-$/+&11500&1&190 &   \\
173650&92036&Si Sr Cr&6.50&4.63(44)&0.09(3)&10140(160)&+/+/+&10000(1410)&2&$-$140 &   \\
175744&92934&Si&6.63&3.22(53)&0.08(1)&12470(320)&+/+/+&12620(140)&2&150 &   \\
176232&93179&Sr&5.90&12.76(29)&0.02&7790(190)&+/+/+&7730(140)&4&$-$60 & 0.01  \\
183806&96178&Cr Eu Sr&5.58&8.22(40)&0.01(1)&9560(30)&+/+/$-$&9940(190)&2&380 &   \\
188041&97871&Sr Cr Eu&5.63&12.47(36)&0&8090(60)&+/+/+&8580(550)&4&490 &   \\
191742&99340&Sr Cr&8.13&3.28(65)&0.06&8290(80)&+/+/+&8300&1&10 &   \\
192678&99672&Cr&7.36&5.05(36)&0&9420(260)&+/+/+&9000&1&$-$420 & 0.09  \\
196502&101260&Sr Cr Eu&5.19&8.25(47)&0&8770(80)&+/+/+&8900(360)&3&130 & 0  \\
201601&104521&Sr Eu&4.69&27.55(62)&0&7740(110)&+/+/+&7780(130)&7&40 & $-$0.13  \\
203932&&Sr Eu&8.81&&0.02&7520(70)&+/+/$-$&7450&1&$-$70 &   \\
204411&105898&Cr&5.30&7.93(24)&0.04&8860(300)&+/+/+&8510(170)&4&$-$350 & $-$0.03  \\
212385&110624&Sr Eu Cr&6.84&7.92(63)&0&8800(530)&+/+/$-$&9200&1&400 &   \\
215441&112247&Si&8.85&0.65(78)&0.21(2)&14780(390)&+/+/+&14000&1&$-$780 &   \\
217522&113711&Sr Eu Cr&7.52&11.36(79)&0.08&6940(10)&+/+/$-$&6750&1&$-$190 &   \\
220825&115738&Cr Sr Eu&4.93&21.25(29)&0&9490(310)&+/+/+&9200(80)&2&$-$290 & $-$0.10(4)  \\
221006&115908&Si&5.65&8.44(29)&0&13330(130)&+/+/$-$&13260(20)&2&$-$70 &   \\
223640&117629&Si Sr Cr&5.17&10.23(31)&0&12210(60)&+/+/+&12240(210)&3&30 &   \\

\hline 
\end{tabular}}
\end{minipage}
\end{table*} 
}

\onltab{6}{
\begin{table*}
\begin{minipage}[t]{180mm}
\caption{The CP3a/b stars.} 
\label{cp3list} 
\centering 
\renewcommand{\footnoterule}{}
\scriptsize{
\begin{tabular}{r r l c r l c l c r l} 
\hline\hline 
HD & HIP & Pec. Type & $m_{v}$  & $\pi$  & $\overline{T}_\mathrm{phot}$  & Phot. & $\overline{T}_\mathrm{eff}$  & \# & $\Delta T$ & $BC$\\
 &  &  & [mag] & [mas] & [K] & &  [K] & & [K] & [mag]\\
\hline 
358&677&Mn Hg&2.08&33.63(35)&13350(100)&+/+/+&13830(40)&2&480 & $-$0.55(15)\\
4335&3604&Hg Mn&6.01&7.63(40)&11690(100)&+/+/+&12000&1&310 & \\
27295&20171&Mn&5.48&12.08(36)&11790(120)&+/+/+&11850(210)&2&60 & \\
27376&20042&Mn Hg&3.54&18.34(15)&12480(160)&+/+/+&12300&1&$-$180 & \\
33904&24305&Hg Mn&3.29&17.54(55)&12530(90)&+/+/+&12160(210)&3&$-$370 & $-$0.64\\
35497&25428&Si Cr Mn&1.65&24.37(33)&13320(170)&+/+/+&13320(100)&2&0 & \\
35548&25365&Hg Mn&6.54&4.64(58)&11060(80)&+/+/+&11500&1&440 & \\
58661&36348&Hg Mn&5.71&6.71(70)&13010(50)&$-$/+/+&13200&1&190 & \\
77350&44405&Sr Cr Hg&5.46&8.31(35)&10490(70)&+/+/+&10250&1&$-$240 & \\
78316&44798&Mn Hg&5.23&6.15(26)&13100(50)&+/+/+&13040(230)&3&$-$60 & $-$0.82\\
89822&50933&Hg Si Sr&4.93&9.61(20)&10600(40)&+/+/+&10950(70)&2&350 & \\
106625&59803&Hg Mn&2.58&21.23(20)&11940(70)&+/+/+&12130&1&190 & \\
143807&78493&Mn Hg&4.97&10.46(24)&11040(70)&+/+/+&10930(460)&2&$-$110 & \\
144206&78592&Mn Hg&4.71&8.76(18)&11940(70)&+/+/+&11740(300)&2&$-$200 & $-$0.68\\
145389&79101&Mn Hg&4.22&15.99(45)&11600(70)&+/+/+&11690(160)&2&90 & \\
147550&80227&Si?&6.22&7.62(64)&10430(110)&+/+/+&10200&1&$-$230 & \\
159082&85826&Hg Mn&6.45&7.39(40)&11200(150)&+/$-$/+&11300&1&100 & \\
190229&98754&Hg Mn&5.65&5.11(32)&12910(100)&+/+/+&13190(440)&2&280 & \\
\hline
4382&3721&Mn P Hg&5.40&4.24(22)&12820(40)&+/+/+&13400&1&580 & \\
19400&14131&He-wk.&5.50&6.34(20)&13530(100)&+/+/+&13000&1&$-$530 & \\
23408&17573&He-wk. Mn&3.88&8.52(28)&12700(90)&+/+/+&11900(990)&2&$-$800 & \\
49606&32753&Mn Hg Si/He-wk.&5.85&3.77(42)&13700(130)&+/+/+&13500&1&$-$200 & \\
51688&33650&Si Mn/He-wk.&6.39&2.72(48)&13380(80)&+/+/+&12500&1&$-$880 & \\
144661&79031&Mn Hg/He-wk.&6.31&8.38(41)&14930(90)&+/+/+&15000&1&$-$70 & $-$1.10\\
144667&79081&He-wk.&6.64&5.85(56)&13350(160)&+/+/+&12900(70)&2&$-$450 & $-$0.91\\
144844&79098&Mn P Ga/He-wk.&5.84&7.35(31)&12600(310)&+/+/+&12430(460)&2&$-$170 & $-$0.77(8)\\
202671&105143&He-wk. Mn&5.38&6.13(31)&13430(260)&+/+/+&13150(70)&2&$-$280 & \\
224926&145&He-wk. Mn&5.10&7.18(30)&13670(40)&+/+/+&14000&1&330 & \\

\hline 
\end{tabular}}
\end{minipage}
\end{table*} 
}

\onltab{7}{
\begin{table*}
\begin{minipage}[t]{180mm}
\caption{The CP4a/ab/b stars.} 
\label{cp4list} 
\centering 
\renewcommand{\footnoterule}{}
\scriptsize{
\begin{tabular}{r r l c r l l c l c r l} 
\hline\hline 
HD/DM & HIP & Pec. Type & $m_{v}$  & $\pi$  & $E(B-V)$  & $\overline{T}_\mathrm{phot}$  & Phot. & $\overline{T}_\mathrm{eff}$  & \# & $\Delta T$ & $BC$ \\
 & & & [mag] &[mas] &  [mag] &  [K] &  &  [K] &  &  [K] & [mag] \\
\hline 
21699&16470&He-wk. Si&5.48&5.38(31)&0.06(1)&15050(100)&+/+/+&16000&1&950 & $-$1.13\\
22470&16803&Si/He-wk.&5.23&6.69(51)&0.01(2)&13630(200)&+/+/+&13760(250)&3&130 & \\
22920&17167&Si/He-wk.&5.52&6.57(48)&0.01(1)&14440(60)&+/+/+&14100(260)&3&$-$340 & $-$0.98\\
28843&21192&He-wk.&5.75&6.86(35)&0.02(2)&14510(240)&+/+/+&14830(150)&3&320 & $-$1.18\\
37058&&He-wk. Sr&7.33&&0.05(2)&18850(200)&+/+/+&19610&1&760 & \\
49333&32504&He-wk. Si&6.06&4.13(51)&0.01(1)&15830(60)&+/+/+&15810(120)&3&$-$20 & \\
62712&37666&He-wk. Si&6.41&5.13(39)&0.01(2)&13430(220)&+/+/+&13530(240)&3&100 & \\
74196&42535&He-wk.&5.55&6.78(26)&0.01(1)&13530(40)&+/+/+&13950(350)&2&420 & \\
79158&45290&He-wk.&5.29&5.60(31)&0.01(1)&13080(160)&+/+/+&13250(70)&2&170 & $-$0.85(14)\\
90264&50847&He-wk.&4.96&8.13(18)&0.02(2)&14230(280)&+/+/+&14600&1&370 & \\
109026&61199&He-wk.&3.83&10.04(13)&0.01(1)&15350(140)&+/+/+&15500&1&150 & \\
137509&76011&Si Cr Fe/He-wk.&6.90&5.12(38)&0.04(1)&14030(680)&+/+/+&12680(110)&2&$-$1350 & \\
142301&77909&He-wk. Si&5.86&6.31(44)&0.11(2)&16100(200)&+/+/+&15860(150)&3&$-$240 & $-$1.36(5)\\
142990&78246&He-wk.&5.41&5.86(24)&0.10(2)&17040(120)&+/+/+&17700(1130)&2&660 & \\
143699&78655&He-wk.&4.88&8.17(30)&0.02(1)&15100(250)&+/+/+&15490(410)&2&390 & $-$1.32\\
144334&78877&He-wk.&5.90&6.21(66)&0.09(1)&15350(70)&+/+/+&14760(380)&3&$-$590 & $-$1.26(8)\\
146001&79622&He-wk.&6.04&6.73(40)&0.17(1)&13510(90)&+/+/+&13790(300)&2&280 & \\
162374&87460&He-wk.&5.87&3.85(46)&0.08(1)&16210(260)&+/+/+&15900(280)&2&$-$310 & $-$1.59\\
175362&92989&He-wk. Si&5.36&7.57(27)&0.04(2)&16890(330)&+/+/+&16790(620)&3&$-$100 & $-$1.43\\
217833&113797&He-wk.&6.50&3.84(57)&0.08(3)&15150(300)&+/+/+&15450&1&300 & \\
\hline
5737&4577& He-wk. & 4.30&4.19(18)&0.01(1)&13790(150)&+/+/+&13970(400)&3&180 & $-$0.83\\
125823&70300& He-wk. & 4.40&7.14(17)&0.02(1)&18240(420)&+/+/+&18940(500)&3&700 & $-$1.74(5)\\
\hline
$-$27 3748&34781&He-rich&9.24&$-$0.17(112)&0.07(2)&21880(280)&$-$/+/+&23000&1&1120 & \\
$-$46 4639&&He-rich&10.02&&0.36&22850&$-$/+/$-$&22500&1&$-$350 & \\
$-$62 2124&&He-rich&11.03&&0.33(5)&25940&+/$-$/+&26000&1&60 & \\
36485&25930&He-rich&6.81&4.72(58)&0.05(1)&17990(310)&+/+/+&18000&1&10 & \\
37017&26233&He-rich&6.55&2.63(73)&0.07(2)&18970(350)&+/+/+&18950(640)&2&$-$20 & \\
37479&&He-rich&6.67&4.84(71)&0.07(1)&21590(310)&+/+/+&22500(710)&2&910 & \\
37776&26742&He-rich&6.99&3.04(55)&0.09(2)&21870(570)&+/+/+&22270(640)&3&400 & \\
58260&35830&He-rich&6.73&2.44(32)&0.09(1)&19030(140)&+/+/+&19000(0)&3&$-$30 & \\
60344&36707&He-rich&7.73&0.25(61)&0.06(2)&21010(260)&+/+/+&22500(2120)&2&1490 & \\
64740&38500&He-rich&4.61&4.29(15)&0.02(1)&22740(200)&+/+/+&22270(1100)&3&$-$470 & \\
66522&39246&He-rich&7.19&2.27(36)&0.27(1)&19210(90)&+/+/+&18000&1&$-$1210 & \\
92938&52370&He-rich&4.79&7.19(20)&0.03(1)&14940(130)&+/+/+&15000&1&60 & \\
96446&54266&He-rich&6.69&2.13(45)&0.09(2)&21620(240)&+/+/+&20950(640)&2&$-$670 & \\
108483&60823&\footnote{He-rich according to \citet{ZN97}.}&3.90&7.92(18)&0.02&18710(240)&+/+/+&19200&1&490 & \\
133518&73966&He-rich&6.38&2.24(44)&0.11(2)&18700(260)&+/+/+&18250(1060)&2&$-$450 & \\
260858&&He-rich&9.14&&0.31&18200&$-$/+/$-$&18000&1&$-$200 & \\
264111&32581&He-rich&9.64&1.26(37)&0.28(1)&22330(260)&+/+/+&21700(990)&2&$-$630 & \\
\hline 
\end{tabular}}
\end{minipage}
\end{table*} 
}

\onllongtab{8}{

\scriptsize
\begin{longtable}{r r l r r l}

\caption{References of the compiled effective temperatures for CP stars. The different groups of peculiar stars discussed in the text are separated by horizontal lines.} \\

\hline
\hline 
HD/DM & $\overline{T}_\mathrm{eff}$ [K]& Ref. & HD/DM & $\overline{T}_\mathrm{eff}$ [K] & Ref. \\
\hline 
\endfirsthead
\caption{Continued.} \\
\hline
\hline
HD/DM & $\overline{T}_\mathrm{eff}$ [K] & Ref. & HD/DM & $\overline{T}_\mathrm{eff}$  [K] & Ref. \\
\hline
\endhead
\hline
\endfoot
\hline
\endlastfoot
\centering
\label{cpref}
27628&7210(90) & (1) a; (2) d$_{+}$&171247&12170& (21) d$_{+}$ \\
27749&7340(180) & (1) a; (2),(5) d$_{+}$; (3),(4) d&171782&11500& (26) c$_{+}$\\
27962&9070(210) & (2) d$_{+}$; (6) d; (7) bd$_{+}$&173650&10000(1410) & (18) b; (25) d$_{+}$\\
28226&7450& (2) d$_{+}$&175744&12620(140) & (23),(24) a\\
28355&7950& (2) d$_{+}$&176232&7730(140) & (18) b; (28) bd$_{+}$; (38) a; (51) b$_{+}$\\
28527&8000(100) & (1) a; (2) d$_{+}$&183806&9940(190) & (22) b$_{+}$; (52) bd\\
28546&7640(140) & (1) a; (2) d$_{+}$&188041&8580(550) & (3) d; (18) b; (22) b$_{+}$\\
29140&7940& (2) d$_{+}$&191742&8300& (53) c$_{+}$\\
29499&7690& (2) d$_{+}$&192678&9000& (27) e\\
30210&8100& (2) d$_{+}$&196502&8900(360) & (25) d$_{+}$; (26) c$_{+}$; (28) bd$_{+}$\\
33204&7650& (2) d$_{+}$&201601&7780(130) & (3) d; (18) b; (20),(28) bd$_{+}$ \\
33254&7760(190) & (1) a; (2) d$_{+}$&&& (22) b$_{+}$; (37) a; (50) d$_{+}$\\
58142&9500& (7) bd$_{+}$&203932&7450& (54) d$_{+}$\\
67523&6700& (8) a&204411&8510(170) & (20),(28),(55) bd$_{+}$; (38) a\\
78362&7220(230) & (2),(5) d$_{+}$&212385&9200& (22) b$_{+}$\\
94334&10030& (9) d$_{+}$&215441&14000& (26) c$_{+}$\\
95418&9600(10) & (10) b; (11) bd$_{+}$&217522&6750& (50) d$_{+}$\\
95608&8950(430) & (2) d$_{+}$; (12) bd&220825&9200(80) & (32),(38) a\\
97633&9250(180) & (10) b; (13) a&221006&13260(20) & (23),(24) a\\
141795&8420& (14) bd&223640&12240(210) & (18) b; (23),(24) a\\
162132&8800& (15) d$_{+}$& \multicolumn{2}{c}{\rule{80pt}{0.1pt}} &\\
173648&8160& (12) bd&358&13830(40) & (56) cd$_{+}$; (70) bd$_{+}$\\
182564&9130& (11) bd$_{+}$&4335&12000& (3) d\\
188728&9530& (9) d$_{+}$&27295&11850(210) & (3) d; (28) bd\\
189849&7820(110) & (2) d$_{+}$; (16) d; (17) bd$_{+}$&27376&12300& (42) a\\
196724&10200& (14) bd&33904&12160(210) & (3) d; (28) bd; (32) a\\
206088&7640& (18) b&35497&13320(100) & (10) b; (28) bd\\
209625&7820(100) & (2) d$_{+}$; (3) d; (17) bd$_{+}$&35548&11500& (3) d\\
213320&10130& (14) bd&58661&13200& (3) d\\
214994&9590(20) & (3) d; (10) b; (19) bd&77350&10250& (28) bd\\
\multicolumn{2}{c}{\rule{80pt}{0.1pt}} &  &78316&13040(230) & (23) a; (28) bd; (58) c$_{+}$\\
8441&9200& (20) bd$_{+}$&89822&10950(70) & (25) d$_{+}$; (59) bd$_{+}$\\
9484&10200& (21) d$_{+}$&106625&12130& (28) bd\\
12098&7800& (22) b$_{+}$&143807&10930(460) & (3) d; (60) bd\\
12767&13000(340) & (21) d$_{+}$; (23),(24) a&144206&11740(300) & (28) bd; (61) b$_{+}$\\
15089&8280(40) & (25) d$_{+}$; (26) c$_{+}$&145389&11690(160) & (3) d; (28) bd\\
18610&8100& (22) b$_{+}$&147550&10200& (28) bd\\
19832&12430(360) & (26) c$_{+}$; (27) e; (28) bd$_{+}$&159082&11300& (15) d$_{+}$\\
23387&8250& (26) c$_{+}$&190229&13190(440) & (3) d; (28) bd\\
24155&13780(70) & (23),(24) a &  \multicolumn{2}{c}{\rule{80pt}{0.1pt}}& \\
24712&7290(60) & (29) d; (30) d$_{+}$&4382&13400& (3) d\\
25823&12720(200) & (26) c$_{+}$; (27) e; (31) d&19400&13000& (62) e\\
26571&11750& (26) c$_{+}$&23408&11900(990) & (62) e, (63) bd\\
27309&11930(250) & (26) c$_{+}$; (27) e&49606&13500& (62) e\\
32549&9730& (32) a&51688&12500& (62) e\\
34452&13830(810) & (26) c$_{+}$; (27) e&144661&15000& (62) e\\
37470&13000& (26) c$_{+}$&144667&12900(70) & (64),(65) d$_{+}$\\
37808&12890(200) & (23),(24) a; (33) d$_{+}$&144844&12430(460) & (26) c$_{+}$; (62) e\\
40312&10180(200) & (26) c$_{+}$; (27) e; (32) a; (34) bd$_{+}$&202671&13150(70) & (62) e; (66) d$_{+}$\\
43819&10930(290) & (20) bd$_{+}$; (23),(24) a; (26) c$_{+}$&224926&14000& (42) a\\
60435&8100& (22) b$_{+}$&\multicolumn{2}{c}{\rule{80pt}{0.1pt}} & \\
62140&7800(140) & (22) b$_{+}$; (25) d$_{+}$&21699&16000& (67) d$_{+}$\\
65339&8170(150) & (25) d$_{+}$; (26) c$_{+}$; (27) e&22470&13760(250) & (23),(42) a; (62) e\\
71866&9000& (25) d$_{+}$&22920&14100(260) & (21) d$_{+}$; (42) a; (62) e\\
75445&7700& (22) b$_{+}$&28843&14830(150) & (23),(42) a; (62) e\\
81009&8250(350) & (25) d$_{+}$; (35) b&37058&19610& (23) a\\
90569&10500& (25) d$_{+}$&49333&15810(120) & (23),(42) a; (62) e \\
92664&14300(90) & (23),(24) a; (26) c$_{+}$&62712&13530(240) & (23),(24),(42) a\\
94427&7500& (25) d$_{+}$&74196&13950(350) & (42) a; (62) e\\
108662&10330(570) & (18) b; (25) d$_{+}$; (26) c$_{+}$; (28) bd$_{+}$&79158&13250(70)& (27) e; (68) cd$_{+}$\\
108945&8700(240) & (18) b; (25) d$_{+}$; (26) c$_{+}$; (27) e&90264&14600& (42) a\\
&& (28) bd$_{+}$; (36) a c$_{+}$&109026&15500&(42) a\\
110066&9030(60) & (20) bd$_{+}$; (22) b$_{+}$; (25) d$_{+}$&137509&12680(110) & (42) a; (69) d$_{+}$\\
111133&9850(220) & (18) b; (25) d$_{+}$&142301&15860(150) & (23),(42) a; (62) e\\
112185&9350(290) & (25) d$_{+}$; (28) bd$_{+}$; (32),(37) a&142990&17700(1130) & (42) a; (62) e\\
112413&11210(360) & (26) c$_{+}$; (37),(38) a; (39) b$_{+}$; (40) bd$_{+}$&143699&15490(410) & (23),(42) a\\
115708&7760(350) & (25) d$_{+}$; (29) d&144334&14760(380) & (23),(42) a; (62) e\\
116114&7850(210) & (22) b$_{+}$; (25) d$_{+}$&146001&13790(300) & (23),(42) a\\
118022&9460(410) & (18) b; (25) d$_{+}$; (26) c$_{+}$; (28) bd$_{+}$&162374&15900(280) & (42) a; (62) e\\
&& (32) a; (41) ac$_{+}$&175362&16790(620) & (23),(24) a; (62) e\\
120198&10080(330) & (26) c$_{+}$; (27) e; (28) bd$_{+}$&217833&15450& (67) d$_{+}$\\
124224&12540(300) & (23),(24),(32),(42) a; (26) c$_{+}$; (28) bd$_{+}$&\multicolumn{2}{c}{\rule{80pt}{0.1pt}}& \\
125248&9650(260) & (25) d$_{+}$; (26) c$_{+}$; (27) e; (41) a c$_{+}$&5737&13970(400) & (42) a; (66) d$_{+}$; (70) b$_{+}$\\
126515&9500& (25) d$_{+}$&125823&18940(500) & (23),(42) a; (62) e\\
128898&7660(340) & (43) d$_{+}$; (44) e&\multicolumn{2}{c}{\rule{80pt}{0.1pt}}& \\
133029&10880(300) & (26) c$_{+}$; (27) e; (28),(45) bd$_{+}$&$-$27 3748&23000& (71) d$_{+}$\\
133792&9300(140) & (46),(47) d$_{+}$&$-$46 4639&22500& (71) d$_{+}$\\
133880&10700(60) & (23),(24) a&$-$62 2124&26000& (71) d$_{+}$\\
134305&8200& (25) d$_{+}$&36485&18000& (71) d$_{+}$\\
137909&8340(360) & (18) b; (22) b$_{+}$; (25) d$_{+}$; (28) bd$_{+}$&37017&18950(640) & (42) a; (71) d$_{+}$\\
137949&7530(40) & (22) b$_{+}$; (25) d$_{+}$&37479&22500(710) & (62) e; (71) d$_{+}$\\
140160&9120(60) & (18) b; (28) bd$_{+}$; (38) a&37776&22270(640) & (42) a; (62) e; (71) d$_{+}$\\
144897&11250& (48) d$_{+}$&58260&19000(0) & (42) a; (62) e; (71) d$_{+}$\\
148112&9220(160) & (18) b; (38) a&60344&22500(2120) & (62) e; (71) d$_{+}$\\
149822&10750& (28) bd$_{+}$&64740&22270(1100) & (42) a; (62) e; (71) d$_{+}$\\
151525&9240(130) & (18) b; (38) a&66522&18000& (71) d$_{+}$\\
152107&8730(230) & (25) d$_{+}$; (26) c$_{+}$; (41) ac$_{+}$&92938&15000& (71) d$_{+}$\\
153882&9450(580) & (18) b; (25) d$_{+}$; (28) bd$_{+}$&96446&20950(640) & (42) a; (71) d$_{+}$\\
155102&9000& (15) d$_{+}$&108483&19200& (71) d$_{+}$\\
157751&11300& (49) d$_{+}$&133518&18250(1060) & (42) a; (71) d$_{+}$\\
166473&7850(210) & (22) b$_{+}$; (50) d$_{+}$&260858&18000& (71) d$_{+}$\\
168733&13320& (23) a&264111&21700(990) & (62) e; (71) d$_{+}$\\
170973&10740(20) & (10) b; (28) bd$_{+}$&&&

\end{longtable}
\noindent
(1) \citealt{S93}; (2) \citealt{SD93}; (3) \citealt{A77}; (4) \citealt{CAY91}; (5) \citealt{vVM96}; (6) \citealt{LS83}; (7) \citealt{AD94b}; (8) \citealt{RM05}; (9) \citealt{CA97}; (10) \citealt{MM85}; (11) 
\citealt{AD96}; (12) \citealt{ACC99}; (13) \citealt{GLU85}; (14) \citealt{AA98}; (15) \citealt{CAT06}; (16) \citealt{TAK84}; (17) \citealt{ACK97}; (18) \citealt{W67}; (19) \citealt{AD88}; (20) \citealt{AP95}; (21) \citealt{LM96}; (22) \citealt{RNW04}; (23) \citealt{L85}; (24) \citealt{M88}; (25) \citealt{BAB94}; (26) \citealt{LS08}; (27) \citealt{SD89}; (28) \citealt{AR00}; (29) \citealt{W97}; (30) \citealt{RLG97}; (31) \citealt{BKD87}; (32) \citealt{GLU87}; (33) \citealt{LCM93}; (34) \citealt{RHA84}; (35) \citealt{WD00}; (36) \citealt{MM92}; (37) \citealt{SB79}; (38) \citealt{SB85}; (39) \citealt{RPS99}; (40) \citealt{KPI02}; (41) \citealt{MO92}; (42) \citealt{HG99}; (43) \citealt{KRW96}; (44) \citealt{BNC08}; (45) \citealt{LA99}; (46) \citealt{RLK04}; (47) \citealt{KTR06}; (48) \citealt{RR06}; (49) \citealt{HN07}; (50) \citealt{G98}; (51) \citealt{RS00}; (52) \citealt{KW78}; (53) \citealt{KS99}; (54) \citealt{GKW97}; (55) \citealt{RLK05}; (56) \citealt{D82}; (57) \citealt{RMA99}; (58) \citealt{ZM87}; (59) \citealt{AD94a}; (60) \citealt{AD89}; (61) \citealt{ZAH07}; (62) \citealt{CAT07}; (63) \citealt{MHS81}; (64) \citealt{CLD04}; (65) \citealt{CH07}; (66) \citealt{LM97}; (67) \citealt{GLC06}; (68) \citealt{WSB06}; (69) \citealt{K06}; (70) \citealt{LAP01}; (71) \citealt{ZN97}
\\ \\
a: Infrared Flux Method; b: fitting models to visual energy distribution; c: fitting models to total energy distribution; d: (Balmer) line profile fitting; e: methods as described in Sect. \ref{collection}. The + sign indicates that models different to solar ones were used or the solar model is justified according to abundance analysis or tests as described in the respective reference. References using the same method are combined.
}

\begin{acknowledgements}
We are grateful to the referee K. St\c{e}pie{\'n} for valuable comments which helped to improve the paper. This research was performed within the project {\sl P17920} of the Austrian Fonds zur F{\"o}rderung der 
wissen\-schaft\-lichen Forschung (FwF) and has made use of the SIMBAD database, operated at CDS, Strasbourg, France.

\end{acknowledgements}

\end{document}